\documentclass[aps,amsmath,amssymb,twocolumn,prl,longbibliography]{revtex4-1}
\usepackage{bm}
\usepackage{epsfig}
\usepackage{amsmath}
\usepackage{color}
\newcommand{\nix}[1]{}
\usepackage[unicode=true,colorlinks=true,urlcolor=blue,citecolor=blue]{hyperref}

\begin{document}

\title{Terahertz Spin-Light Coupling in Proximitized Dirac Materials}

\author{Konstantin S. Denisov$^{1,2}$}
\author{Igor V. Rozhansky$^{1}$}
\author{Sergio O. Valenzuela$^{3,4}$}
\author{Igor {\v{Z}}uti{\'c}$^{2}$}

\affiliation{$^{1}$Ioffe Institute, 194021 St. Petersburg, Russia}
\affiliation{$^{2}$Department of Physics, University at Buffalo, State University of New York, Buffalo, NY 14260, USA}
\email{denisokonstantin@gmail.com}
\affiliation{$^{3}$Catalan Institute of Nanoscience and Nanotechnology (ICN2), CSIC and BIST,
Barcelona, Spain}
\affiliation{$^{4}$Instituci\'{o} Catalana de Recerca i Estudis Avan\c{c}ats (ICREA)
Barcelona 08010, Spain}

%\affiliation{$^{2}$Department of Physics, University at Buffalo, State University of New York, Buffalo, NY 14260, USA}

\begin{abstract}
The %IZ atomically-thin 
two-dimensional (2D) materials are highly 
susceptible to the influence of their neighbors, thereby enabling the design by proximity phenomena. %IZ Here, 
We reveal a remarkable terahertz (THz) spin-light interaction in 2D Dirac materials that arises from magnetic and spin-orbital proximity effects. The dynamical realization of the spin-charge conversion, the electric dipole spin resonance (EDSR), of Dirac electrons displays distinctive THz features, %IZ in the THz range, 
upon emerging spin-pseudospin proximity terms in the Hamiltonian. To capture the effect of fast pseudospin dynamics on the electron spin, we
develop a mean-field theory and complement it with a quantum-mechanical treatment. As a specific example, we investigate the THz response of a single graphene layer proximitized by a magnetic substrate. Our analysis demonstrates a strong enhancement and anomalous polarization structure of the THz-light absorption which can enable
%IZ offers promising prospects for 
THz detection and efficient generation and control of spins in spin-based quantum devices. The identified coupled spin-pseudospin dynamics is not limited to EDSR and may influence a broad
range of optical, transport, and ultrafast phenomena.  %IZ and transport properties.	
\end{abstract}	
\date{\today}
\maketitle

Heterostructures combining two-dimensional (2D) van der Waals (vdW)
materials offer innovative approaches for tailoring material 
properties~\cite{geim2013van,novoselov20162d}. The atomically-thin 2D layers imply that many phenomena can be dominated by proximity effects~\cite{vzutic2019proximitized,sierra2021van}. 
This scenario is exemplified in spin-dependent properties of graphene-vdW heterostructures~\cite{vzutic2019proximitized,sierra2021van,han2014graphene,han2016perspectives,avsar2020colloquium}.  Transition metal dichalcogenides  
imprint spin-pseudospin-valley splitting and spin-orbit coupling (SOC) onto graphene~\cite{sierra2021van,gmitra2015graphene,gmitra2016graphene,garcia2018spin,david2019induced}, leading to spin filtering~\cite{cummings2017anisotropy,benitez2018anisotropy,ghiasi2017anisotropy} as well as enhanced spin-to-charge 
interconversion~\cite{offidani2017optimal,garcia2017SHE,safeer2019MoS2,benitez2020SHEISGE,galceran2021CSI}. 
An exchange field and ensuing carrier spin splitting can be further induced in graphene through magnetic proximity~\cite{yang2013proximity,lazic2016effective,asshoff2017magnetoresistance,wei2016strong,wu2020large,ghiasi2021electrical}. 

These added functionalities can pave the way to novel topological phases and devices that merge spin injection, detection, and manipulation into a single graphene platform~\cite{sierra2021van,avsar2020colloquium,Wen2016:PRA}. Graphene and other Dirac materials have a great potential for THz (opto)electronics. Fast, room-temperature THz detectors made of graphene exhibit excellent sensitivity, high dynamic range, and broadband operation~\cite{castilla2019fast}. Massless Dirac fermions in graphene and topological insulators have large nonlinear optical coefficients and harmonic conversion efficiencies, suitable for THz high-power harmonic generation~\cite{hafez2020THz,kovalev2021THz,tielrooij2021THz}.

\begin{figure}[htbp]
	\centering
	\includegraphics[width=0.5\textwidth]{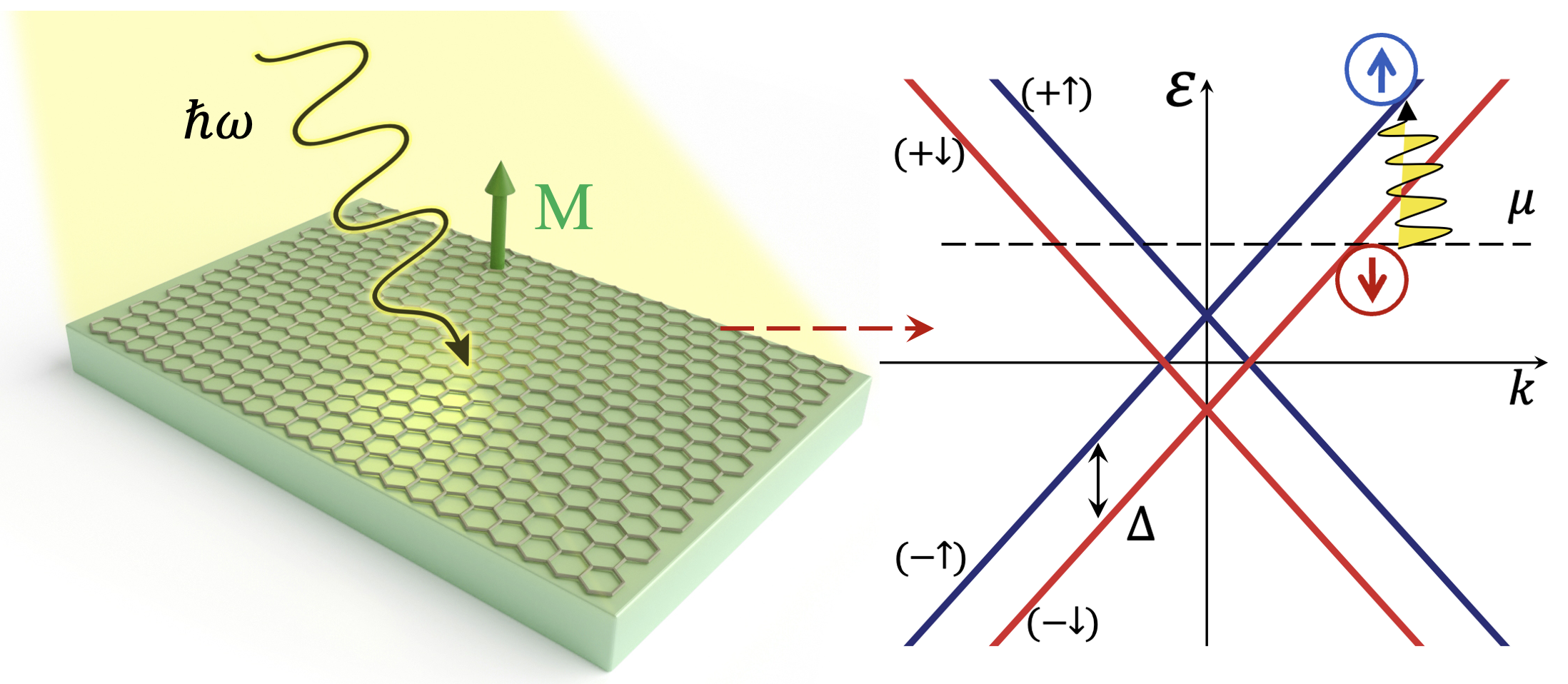}
	\caption{Electric field of THz radiation causes intersubband spin-flip transitions in a graphene on a substrate with a magnetization, {\bf M}. The Dirac spectrum with a proximity-induced spin splitting, $\bm{\Delta}$, wave vector, {\bf k}, and the Fermi energy, $\mu$.
	}
\label{fig:1}
\end{figure}

In this work, we investigate the spin-charge THz
dynamics in proximitized 2D Dirac material  
with spin splittings, as shown in Fig.~\ref{fig:1}, and describe the resulting spin-light interaction including SOC. 
Our results demonstrate an unexplored realization of the spin-charge conversion from the electric dipole spin resonance (EDSR)~\cite{rashba1960properties,Rashba2002:JS}, the excitation of electron spin precession by an ac electric field, which is a versatile tool from probing SOC, inhomogeneous magnetism, and topological states to realizing spin injection and controlling qubits~\cite{Bell1962:PRL,McCombe1967:PRL,Pekar1964:SPJETP,Zutic2004:RMP,duckheim2006electric, wilamowski2008elec,nadj-perge2010qubit,Stier2023:PRB,Brooks2020:PRB,Wang2022:NC,Stano2022:NRM}.
In the presence of SOC, the EDSR is driven by a unique mechanism due to coupled spin-pseudospin dynamics~\cite{tuan2014pseudospin,demoraes2020entanglement}. 
Previously overlooked, this phenomenon becomes crucial at frequencies $\omega$ in the THz range, where $\omega\tau_p\gg1$, with $\tau_p$ the momentum relaxation time. 
We calculate the absorption in proximitized graphene, using realistic SOC and magnetic exchange parameters, 
and demonstrate that the predicted EDSR leads to a remarkable increase of both the spin susceptibility and THz absorption. 
We reveal an anomalous polarization structure of EDSR
controlled by the coupled spin-pseudospin dynamics and transformed 
for massive Dirac electrons upon lowering 
the Fermi energy, 
$\mu$. Our findings provide: (i) striking differences 
from prior mechanisms~\cite{duckheim2006electric,duckheim2007resonant,wilamowski2007g,wilamowski2008elec} and
(ii) highlight their relevance 
for THz detection and spin manipulation.

The electron spin resonance (ESR)~\cite{Zutic2004:RMP,weil2007electron} is a well-established technique for studying spin phenomena in solids. 
It requires a static magnetic field that defines 
the direction of an equilibrium spin polarization, ac-magnetic field, that induces spin-flip transitions, 
which are detected by the absorption 
close to the Larmor frequency. 
The EDSR is essentially identical, but with 
the spin-flip transitions induced by ac-electric field, allowed in the presence of 
SOC~\cite{rashba1960properties,Rashba2002:JS,rashba2003efficient}. 
In nanostructures the SOC symmetry and magnitude can be designed 
to increase the efficiency of spin-flip absorption 
for a stronger spin-light interaction in the EDSR than in ESR. 
This is very desirable for 
semiconductor qubits~\cite{nadj-perge2010qubit,Wang2022:NC,Stano2022:NRM}
and to resonantly enhance the spin-charge conversion.

We analyze the EDSR for the low-energy 
Hamiltonian of 
Dirac electrons in a hexagonal system
near the $K$ and $K'$ valleys, including magnetic-exchange and SOC 
\begin{equation}
\label{eq:Htot}
    \mathcal{H} = \mathcal{H}_0 + \mathcal{H}_{\rm ex} + \mathcal{H}_{\rm so},
\end{equation}
where $\mathcal{H}_0 = \hbar \bm{\Omega}_k \cdot \bm{\tau}$ 
defines 
the Dirac spectrum,  $\hbar$ is the Planck constant, 
with
$\bm{\tau}$ the lattice pseudospin operator~\cite{Kim2017}, 
$\bm{\Omega}_k = 2 v_F (\xi  k_x, k_y, U/2 \hbar v_F)$ 
the Larmor frequency,
$\bm{k}$ an electron wave vector, 
$v_F$ the Fermi velocity, 
and $\xi = \pm 1$ the valley index.
In $z$ component of $\bm{\Omega}_k$, $U$ 
is the strength of the staggered potential, due to 
the on-site asymmetry between two inequivalent sublattices in 2D hexagonal lattices
due to different atoms in unit cell, or 
from the effect of a substrate. 
$\mathcal{H}_{\rm ex} = \bm{\Delta} \cdot \bm{s}$
describes the magnetic exchange, where $\bm s$ is the spin operator,
and $\Delta$ the spin splitting 
in the meV (THz)  
range \cite{sierra2021van}, whereas 
$\mathcal{H}_{\rm so} = \hbar \bm{\Omega}_{\rm so}(\bm{\tau}) \cdot \bm{s}$ characterizes the SOC, where  we assume  
a $\bm{k}$-independent $\bm{\Omega}_{\rm so}(\bm{\tau})$.  
For a graphene/TMD, $\mathcal{H}_{\rm ex}$ is %IZ replaced by 
the valley-dependent splitting $\xi \bm{\Delta} \cdot \bm{s}$.

A hallmark of 
Dirac materials 
is the spin-pseudospin coupling
and their entanglement~\cite{garcia2018spin,sierra2021van,tuan2014pseudospin,demoraes2020entanglement} arising from $\mathcal{H}_{\rm so}$.  
To model the spin-pseudospin dynamics driven by an electromagnetic wave, we consider the interaction $\mathcal{V} = \hbar \bm{\Omega}_{\rm int}(t) \cdot \bm{\tau}$, 
where $\bm{\Omega}_{\rm int} = - 2 (e/ \hbar c) v_F (\xi A_x, A_y)$, with $\bm{A}$ the vector potential.
Assuming normal incidence, we focus on the spin-light coupling emerging via an
electric field component $\bm E_\omega = (i \omega/c) \bm A_\omega$.

\begin{figure}[t]
	\centering
\includegraphics[width=0.4\textwidth]{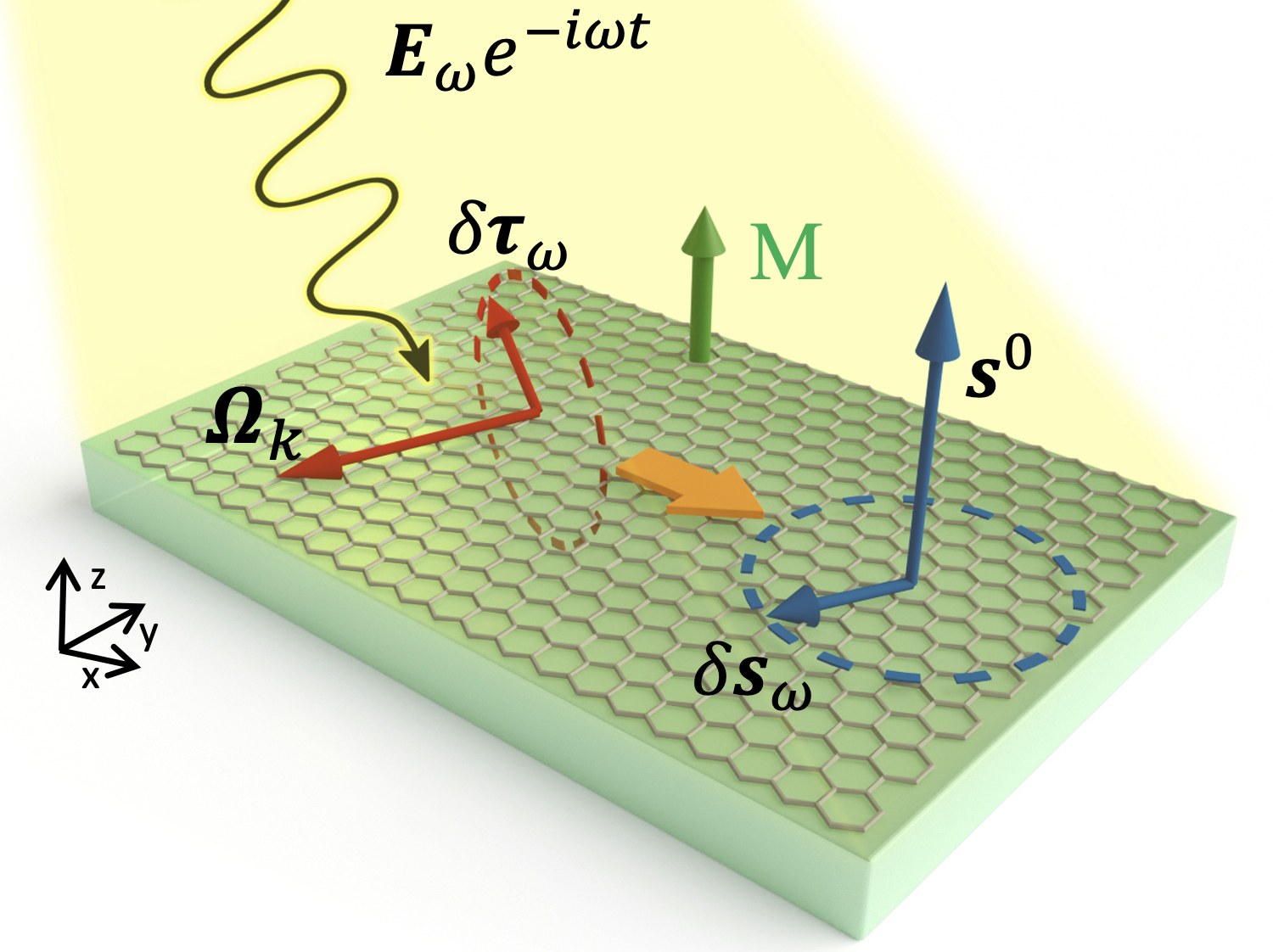}
\caption{EDSR of a Dirac electron driven by coupled spin-pseudospin dynamics.  An incident THz radiation (yellow) with electric field $E_\omega$ is absorbed creating a pseudospin component, $\delta \tau_\omega$, which precesses in the pseudospin field, $\Omega_k$, exerting a torque 
on a spin due to SOC (depicted by the orange arrow). Therefore, a spin component, $\delta s_\omega$, precesses $\perp$  to the spin splitting $\mathbf \Delta$, which is along the equilibrium spin $\mathbf s^0$.   	
}
\label{fig:2}
\end{figure}

The coupled spin-pseudospin dynamics
can be described by the mean-field 
equations of motion for the classical vectors $\bm{\tau}, \bm{s}$~\cite{SM}
\begin{align}
	& 
	\dot{\bm \tau}
	=  \left[
	\left(
	\bm{\Omega}_k +
	\bm{\Omega}_{\rm so}'(\bm{s})
	\right)
	\times \bm{\tau}
	\right]
	+
	 \left[
	\bm{\Omega}_{\rm int}(t) \times \bm{\tau}
	\right],
\label{eq:tau} \\
	&
	\dot{\bm s} =
	\left[
	\left(
        \bm{\Omega}_{\rm ex}
	+
	\bm{\Omega}_{\rm so}(\bm{\tau})
	\right)
	\times \bm{s}
	\right],
	\label{eq:s} 
\end{align}
where $\bm{\Omega}_{\rm so}'(\bm{s})$ is obtained from $\mathcal{H}_{\rm so}$ with $ \bm{\tau}\bm{\Omega}_{\rm so}'(\bm{s}) =\bm{s} \bm{\Omega}_{\rm so}(\bm{\tau})$
and $\hbar \bm{\Omega}_{\rm ex} = \bm{\Delta}$. For $\Omega_{\rm so} \lesssim \Omega_{\rm ex}$,
this model captures the 
spin resonance at $\hbar \omega \approx \Delta$, where  $\bm E_\omega$
induces the dynamics of $\bm{\tau}$, which triggers the {$\bm{s}$} precession 
due to spin-pseudospin coupling \cite{tuan2014pseudospin}, as depicted 
in Fig.~\ref{fig:2}. 
For $\Omega_{\rm ex} \tau_p~\gg~1$, 
the intersubband spin-light coupling results in a resonant {absorption} peak, 
as discussed below. 

We provide our framework for {$n$}-doped graphene with $U=0$; 
the case of $U\neq 0$ 
%for massive Dirac electrons it 
is given in~\cite{SM} (see also Ref.~\cite{nomura2007quantum,barati2017optical} therein). 
The dynamics of 
two-level systems, including 
the lattice pseudospin, can be {modeled} using the classical precession equation for the quantum average of the operator, 
including 
the corresponding Pauli matrices~\cite{sakurai1995modern,griffiths2018introduction}. 	
Ignoring  SOC, Eq.~(\ref{eq:tau})
describes 
the free pseudospin 
oscillations with 
$\Omega_k$ 
and, when subjected to an external oscillating field $\bm{\Omega}_{{\rm int}, \omega} e^{- i \omega t}$,  
captures the resonance at $\omega \approx \Omega_k$, 
related to quantum pseudospin-flip transitions.
An  electron 
from $K$-valley 
with $\bm{k}$ has a static
pseudospin $\bm{\tau}^0 = \bm{\hat}{\bm k}/2$,  parallel to $\bm{\Omega}_k = 2 v_F \bm{k}$. 
Applying $\bm{\Omega}_{\rm int}(t)$ 
generates the torque, $\bm{\Omega}_{\rm int}(t) \times \bm{\tau}^0
\propto \bm{\hat}{\bm k} \times \bm{A}(t)$
which, following 
Eqs.~(\ref{eq:tau}), 
triggers the pseudospin rotation around 
$\bm{\Omega}_k$.
With $\bm{A}(t) = 2 {\rm Re}[ \bm{A}_{\omega} e^{-i\omega t} ]$, 
the linear response 
$\delta \bm{\tau}_\omega= \delta \tau_{z} \bm{\hat}{\bm z}  + 
\delta \tau_{\varphi} \hat{\bm{\varphi}}$ 
is perpendicular to $\bm{\Omega}_k$ 
\begin{equation}
\label{eq:pseudospinRes}
    \delta \tau_{z,\varphi}(\omega) =
    \frac{\alpha_{z,\varphi} \mathcal{T}}{\Omega_k^2 - \omega^2},
    \quad
\mathcal{T} = e\frac{v_F}{\hbar c} [\bm{A}_\omega \times \bm{\hat}{\bm k}]_z,
\end{equation}
where $\alpha_{z,\varphi} = (-i\omega, \Omega_k)$. 
Poles in $\delta \tau_{z,\varphi}(\omega)$ at $\omega = \Omega_k$, 
give the pseudospin resonance i.e. for the
interband transitions leading to 
the universal absorption $\alpha_0= \pi e^2/\hbar c$. 
To obtain the interband absorption $\alpha = (\alpha_0/2) (\Theta(\hbar \omega - 2 \mu + \Delta) + \Theta(\hbar \omega - 2 \mu - \Delta) )$,
$\Theta$ is the Heaviside function, 
one further needs to calculate the real part of the  conductivity
by summing $2 v_F \delta \tau_{\varphi}(\omega+i0)\hat{\bm{\varphi}}_x$
over all quantum numbers with the equilibrium distribution function~\cite{SM}.

To illustrate the spin-pseudospin dynamics, we  consider surface inversion asymmetry 
and  the relevant Bychkov-Rashba SOC~\cite{gmitra2009band,min2006intrinsic} 
\begin{equation}
\label{eq:Hso-Ras}
    \mathcal{H}_{\rm so} = 2\lambda_{\rm so} \left( \xi \tau_x s_y - \tau_y s_x \right),
\end{equation}
where the SOC strength 
$\lambda_{\rm so} \lesssim \Delta$~\cite{vzutic2019proximitized,sierra2021van,asshoff2017magnetoresistance}.
For the spin dynamics, in the lowest order of $\lambda_{\rm so}/\mu$, 
we account for $\mathcal{H}_{\rm so}$ only  in Eq.~(\ref{eq:s}).
The oscillating pseudospin,
$\delta \bm{\tau}(t) = 2 {\rm Re}[ e^{-i\omega t}  \delta \bm{\tau}_\omega]$, 
induced by $\bm{E}_\omega$, 
then contributes to 
$\hbar \bm{\Omega}_{\rm so}(t) = 2\lambda_{\rm so} [\bm{\hat}{\bm z} \times \delta \bm{\tau}(t)]$ and exerts
a torque on $\bm{s}$ (Fig.~\ref{fig:2}). 
For the out-of-plane geometry, $\bm{s}^0 \parallel \bm{\Delta} \parallel \bm{\hat}{\bm z}$,
the resonant spin component 
in Eq.~\ref{eq:s}
and linear in $\lambda_{\rm so}$,
$\delta \bm{s}_\omega = \delta s_k \bm{\hat}{\bm k} + 
\delta s_\varphi \hat{\bm{\varphi}}$, 
is given~by
\begin{equation}
    \label{eq:spinres}
    \delta s_{k,\varphi}(\omega) =
s^0  \beta_{k,\varphi}
\frac{2 \hbar^{-1} \lambda_{\rm so} \Omega_k \mathcal{T}}{\left( \Omega_{\rm ex}^2 - \omega^2\right) \left(\Omega_k^2 - \omega^2\right)},
\end{equation}
where $\beta_{k,\varphi} = (\Omega_{\rm ex}, -i\omega)$
and $s^0 = \pm 1/2$ is the initial spin state.
A pole at $\hbar \omega = \Delta$
corresponds to intersubband 
spin-flip transitions.
This resonance contributes to the absorption, 
which can be calculated from Eq.~(\ref{eq:spinres})  
in the rotating frame~\cite{weil2007electron}
by collecting the spin response from all electrons with 
different quantum numbers~\cite{SM}.

We can complement this analysis by evaluating, quantum mechanically, the EDSR-induced absorption. 
The matrix element $M_{\uparrow \downarrow}$ of the direct intersubband spin-flip transition
from $(+ \downarrow)$ to $(+\uparrow)$ states (see Fig.~\ref{fig:1}) is found from the second-order perturbation theory
\begin{equation}
\label{eq:M-element}
    M_{\uparrow \downarrow} = \frac{\mathcal{V}_{(+\uparrow; -\uparrow)} \mathcal{H}_{(-\uparrow; + \downarrow)}^{\rm so}}{
    \varepsilon_{k,\downarrow}^+
    -\varepsilon_{k,\uparrow}^-
     }
	+
	\frac{
		\mathcal{H}_{(+\uparrow; - \downarrow)}^{\rm so}
		\mathcal{V}_{( -\downarrow; + \downarrow)} }{
		\varepsilon_{k,\downarrow}^+ + \hbar \omega - \varepsilon_{k,\downarrow}^-
		},
\end{equation}
where 
$\varepsilon_{k,s}^\pm = \pm v_F \hbar k + s \Delta$
and $\mathcal{V} = \hbar \bm{\Omega}_{\rm int, \omega} \cdot \bm{\tau}$.
%iz
The spin-generation rate 
is given by 
the Fermi's Golden rule 
\begin{equation}
\label{eq:pump}
W_s = \frac{2\pi}{\hbar} \sum_k 2 | M_{\uparrow \downarrow} |^2 (f_k^{\uparrow} - f_k^{\downarrow})
\mathcal{L}(\hbar \omega),
\end{equation}
where
$f_k^{\uparrow,\downarrow}$ is the Fermi-Dirac  
function of ($\uparrow, \downarrow$)-electrons in
the conduction band,
the factor $2$ accounts for $(K,K')$ valleys, and
the frequency broadening 
$\mathcal{L}(\hbar \omega) = (\gamma/\pi)/[(\hbar \omega - \Delta)^2 + \gamma^2]$ is given by the Lorentzian with the spin-flip 
dephasing rate, $\gamma$. We 
express 
$W_s = \alpha_{\rm sf}(\omega) (I/\hbar \omega)$ in terms of the 
the radiation intensity, $I$, and the absorption coefficient,
$\alpha_{\rm sf}(\omega)$, 
which at zero temperature is  $\alpha_{\rm sf}= \pi \gamma \alpha_{\rm sf}^{\rm max} \mathcal{L}(\hbar \omega)$ with
\begin{equation}
    \label{eq:absorb}
    \alpha_{\rm sf}^{\rm max} 
    = 
    \alpha_0 b \frac{\lambda_{\rm so}^2}{4 \Delta \pi \gamma} \left[
    \ln{\left( \frac{\mu + \Delta}{\mu - \Delta} \right)}
    + \frac{\Delta^3/2\mu}{(\mu^2 - \Delta^2)}
    \right], 
\end{equation}
where 
$b\sim 1$ is a prefactor determined by the directions of $\bm{\Delta}, \bm{E}_\omega$. 
The same expression for $\alpha_{\rm sf}(\omega)$ 
can be obtained using the Kubo formula for the optical conductivity by including SOC in the velocity matrix elements~\cite{SM}.
\begin{figure}[t]
	\centering
	\includegraphics[width=0.5\textwidth]{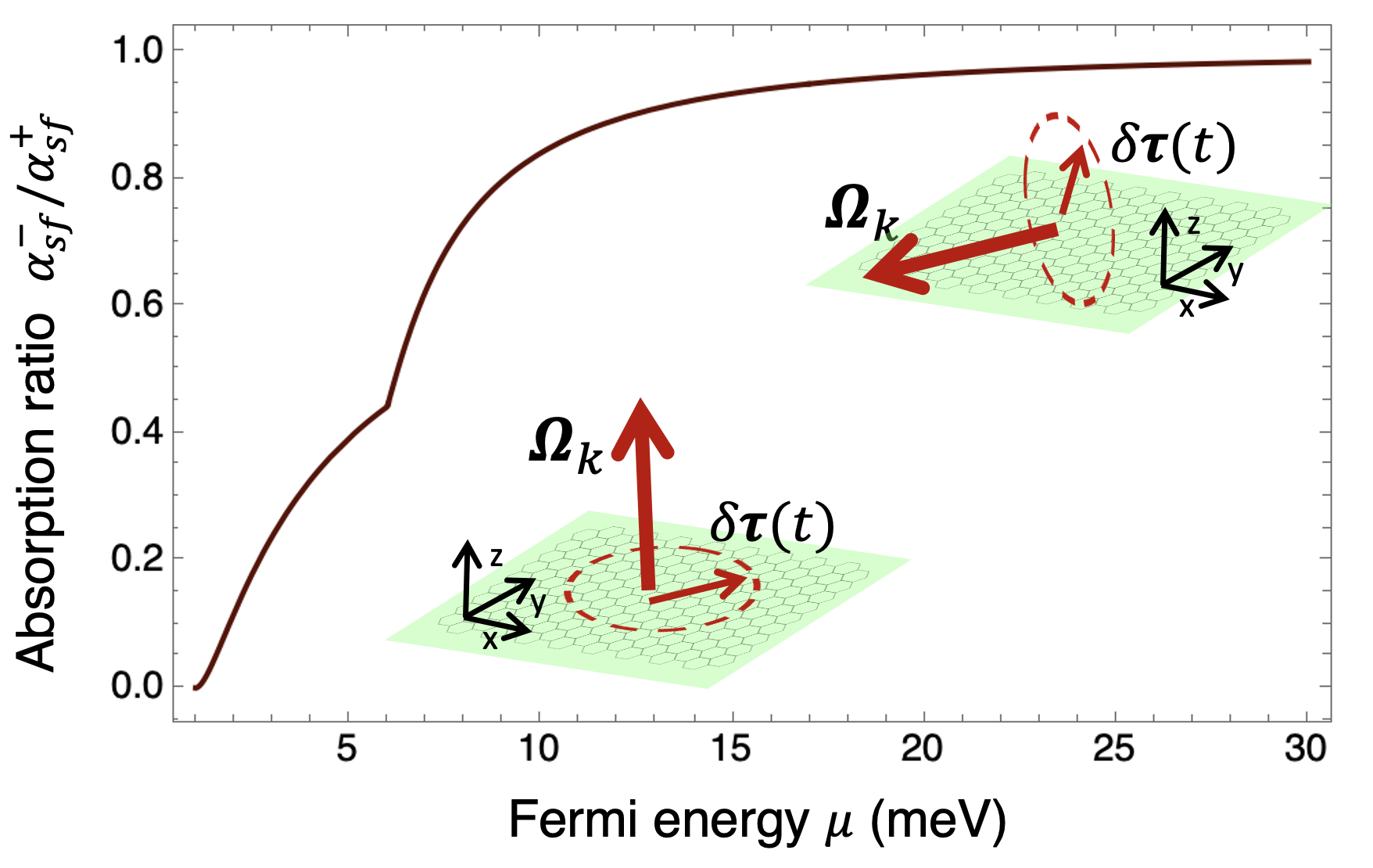}
	\caption{
Ratio of the EDSR absorption for two circular polarizations of $\bm{E}_\omega$ as a function of the Fermi energy. 	The parameters are $\Delta = 5$~meV, $\lambda_{\rm so} = 1.2$~meV, and $U = 3.5$~meV.
	}
	\label{fig:extra1}
\end{figure}

$\alpha_{\rm sf}^{\rm max}$ from Eq.~(\ref{eq:absorb}) has an anomalous and counterintuitive polarization structure encoded in $b$, 
reflecting
the role of both the pseudospin dynamics and SOC field symmetry for EDSR.
Instead of directly interacting with $\bm{E}_\omega$,  a $K$-valley
 electron spin interacts with a SOC field,  
$\hbar \bm{\Omega}_{\rm so}(t) = - 2\lambda_{\rm so}  \delta \tau_{\varphi}(t) \bm{\hat}{\bm k}$ 
linearly polarized, irrespective of $\bm{E}_\omega$. 
For a gapless spectrum, this results in
the suppressed EDSR sensitivity to the $\bm{E}_\omega$ polarization: 
for $\bm{\Delta} \parallel \bm{\hat}{\bm z}$, 
$\alpha_{\rm sf}$
is the same for both circular polarizations with $b=1$,  
while,  in case of the in-plane $\bm{\Delta}$ orientation, 
$b=3/4$ and $b=1/4$, for $\bm{E} \parallel \bm{\Delta}$ and $\bm{E} \perp \bm{\Delta}$, 
respectively. 

\begin{figure}[t]
	\centering
	\includegraphics[width=0.45\textwidth]{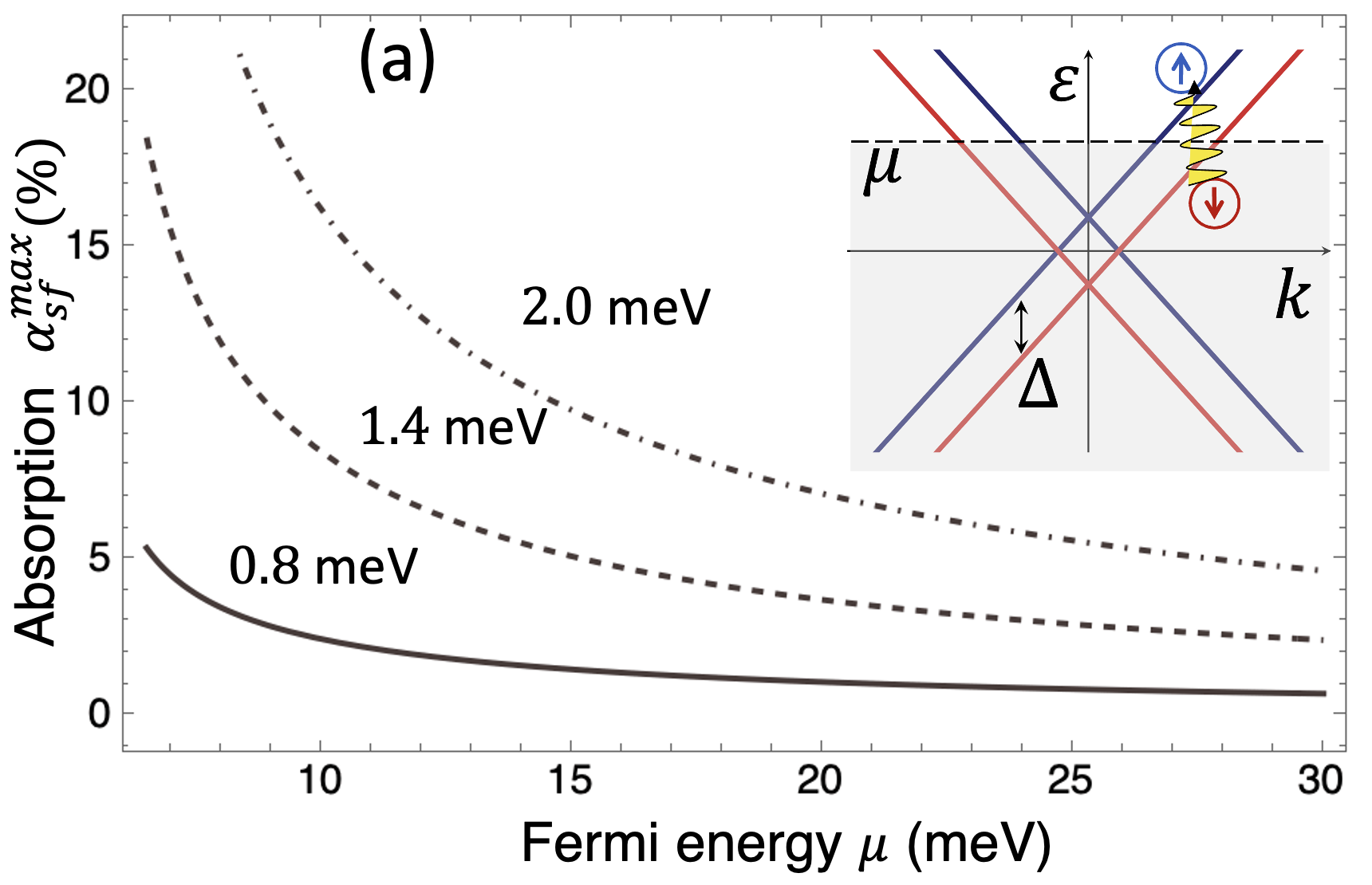}
	\includegraphics[width=0.45\textwidth]{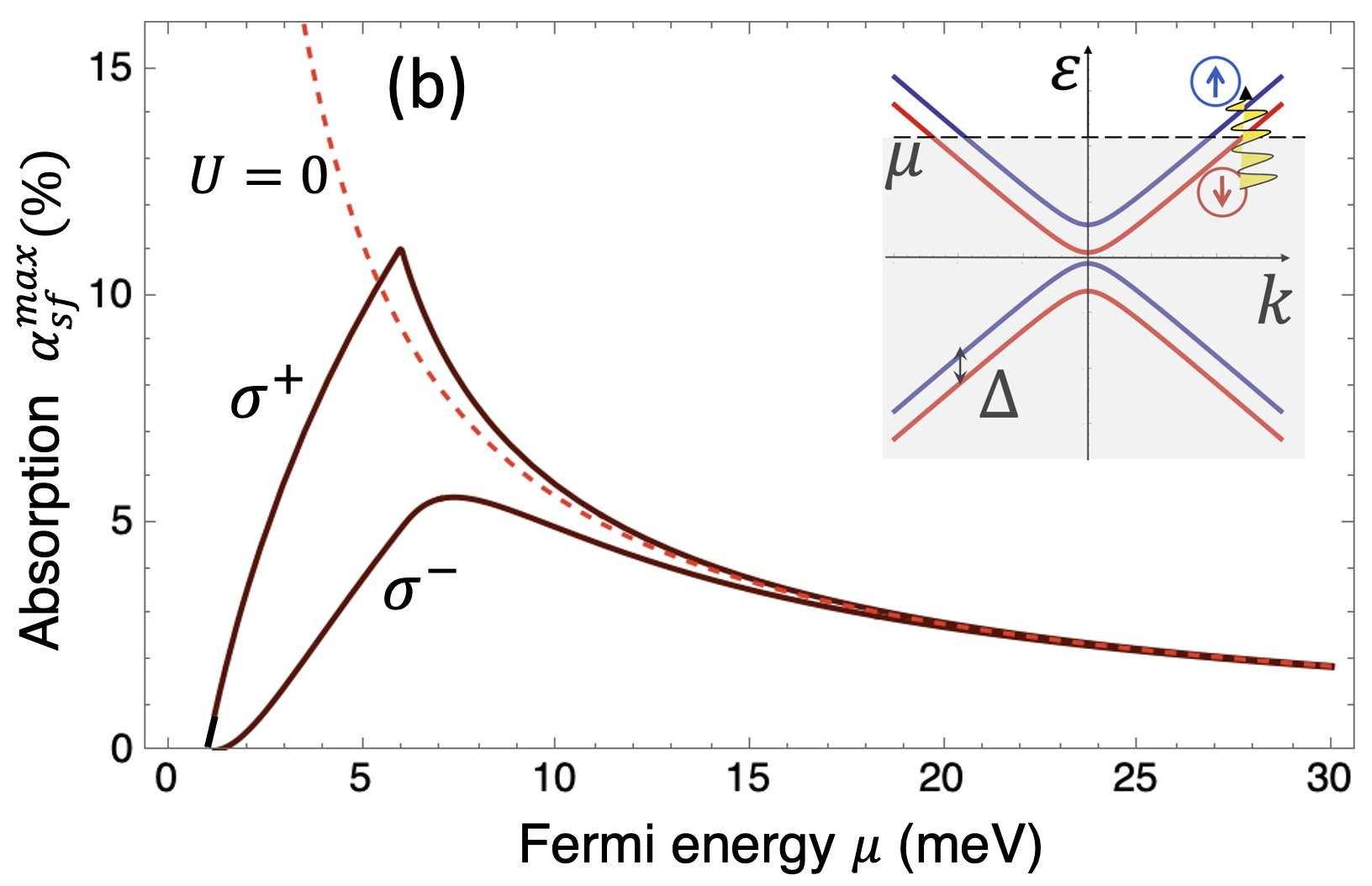}
	\caption{The evolution of the spin-pseudospin coupling-induced EDSR absorption with the Fermi energy and SOC. 
(a): Linear spectrum (inset), the parameters are $\Delta = 5$~meV,	$\hbar/\gamma = 70\,$ps, with $\gamma$ the spin-flip dephasing rate, and $\lambda_{\rm so}= 0.8, 1.4$, and $2.0$~meV. (b): Massive Dirac electrons (inset) for two circular polarizations, $U = 3.5$~meV, $\lambda_{\rm so} = 1.2$~meV. 
	}
	\label{fig:3}
\end{figure}

However, for massive Dirac electrons with 
$\mu \lesssim 2 U$
the EDSR
at $\bm{\Delta} \parallel \hat{\bm{z}}$ and $\Delta>0$ 
is induced preferably for $\sigma^+$ polarization, 
as shown in Fig.~\ref{fig:extra1} from 
$\alpha_{\rm sf}^- / \alpha_{\rm sf}^+$.
In contrast to Eq.~(\ref{eq:pseudospinRes}), 
at $pv_F \ll U$ 
the vector $\bm{\tau}^0 \parallel \hat{\bm{z}}$, and 
$\delta \bm{\tau}_\omega$ lies 
within the electron motion plane, 
implying $\bm{\Omega}_{\rm so}(t) \propto \lambda_{\rm so} 
[\hat{\bm{z}} \times \hat{\bm{\rho}}_\sigma(t) ]$
follows  $\hat{\bm{\rho}}_\sigma(t)$, the unit vector rotating counter- or clockwise depending on $\sigma$. Hence, the spin resonance for $\delta \bm{s}_\omega$
obeys the ordinary polarization rules, i.e. 
at $\bm{\Omega}_{\rm ex} \parallel \hat{\bm{z}}$, 
the EDSR absorption is active {only} for
$\sigma^+$ in both valleys. 
As $\mu$ departs from the conduction band bottom, 
$\bm{\Omega}_k$ and $\bm{\tau}^0$ gradually tilt onto the plane, 
which suppresses the polarization sensitivity of the EDSR 
approaching the result $b=1$ for the linear spectrum at $\mu \gg U$, see Fig.~\ref{fig:3}. 
This {behavior} also contrasts the valley-dependent circular dichroism for the interband spin-conserving absorption~\cite{cao2012valley,mak2012control,xiao2012coupled}, i.e. the fundamental absorption in $K$($K^\prime$) valley occurs for $\sigma^+$($\sigma^-$). A more accurate analysis~\cite{SM} shows that 
at $\mu \lesssim U$ the EDSR does inherit a finite valley dependence 
with slightly different absorptions of $\sigma^+$ light in $K$ and $K^\prime$ valleys.

At smaller frequencies $\,\sim\,$GHz, 
$\omega \tau_p \lesssim 1$, 
there is a change in the mechanism
for the EDSR resonance from Eqs.~(\ref{eq:tau}) and (\ref{eq:s})
to the current-induced spin resonance~\cite{wilamowski2007g,duckheim2006electric}. 
Here, a spin torque acting on 
a 2DEG equilibrium spin density $\bm{S}^0 = \bm{\Delta}/ 2\pi v_F^2 \hbar^2$ 
stems from an effective Larmor frequency $\bm{\Omega}_{\rm so}^{\rm D}(t) = (\lambda_{\rm so}/\hbar v_F) [
\bm{\hat}{\bm z} \times \bm{v}(t)
]$, determined by the Drude velocity
$\bm{v}(t) = 2 {\rm Re}[\bm v_\omega e^{-i\omega t}]$
with $\bm v_\omega =e\bm{E}_\omega \tau (v_F/p_F)/(1 - i \omega \tau)$.
One can qualitatively analyze the emerging nonequilibrium spin density, $\bm{S}_\omega$,
based on the Bloch spin-resonance equation
\begin{equation}
	- i \omega \bm{S}_\omega + 
	T_2^{-1}
	\bm{S}_\omega = \left[ 
	\bm{\Omega}_{\rm ex}
	\times \bm{S}_\omega \right] + \left[ \bm{\Omega}_{{\rm so},\omega}^{\rm D}  \times \bm{S}^0\right].
	\label{eq:Bloch}
\end{equation}
Since $\bm{\Omega}_{{\rm so},\omega}^{\rm D} \propto \left[\bm{\hat}{\bm z} \times \bm{E}_\omega \right]$,  
the resonant 
absorption for $\bm{\Delta} \parallel \bm{\hat}{\bm z}$ 
will be active only for one circular polarization.
For the in-plane geometry, the
absorption only takes place when 
$\bm{E} \parallel \bm{\Delta}$, since
the torque is absent as $ \bm{\Omega}_{{\rm so},\omega}^{\rm D} \parallel \bm{S}^0 $ for $\bm{E} \perp \bm{\Delta}$.
In the intermediate regime, 
$\omega \tau_p \sim 1$, both resonance 
mechanisms (intraband and intersubband) 
should be considered on equal footing.

It is instructive to compare $\alpha_{\rm sf}^{\rm max}$ with $\alpha_0 \approx 2.5 \%$ for graphene. 
For $\mu \gtrsim 2 \Delta$ 
(or $\mu \gtrsim 2 U$ for massive Dirac electrons with $U>\Delta$)
$\alpha_{\rm sf}(\omega) \approx \alpha_0 b ({\lambda_{\rm so}^2}/{2 \mu}) \mathcal{L}(\hbar \omega)$,
with the peak value determined by
$\alpha_{\rm sf}^{\rm max} = \alpha_0 b \lambda_{\rm so}^2/(2\pi \mu \gamma$).
For $\mu = 16$~meV, $\lambda_{\rm so} = 0.7$~meV, and $T_2 = \hbar/\gamma = 70$ ps, we obtain 
$\alpha_{\rm sf}^{\rm max} \approx 0.55 \alpha_0 = 1.25 \%$. 
We also compare the EDSR-induced 
$M_{\uparrow \downarrow}$ from Eq.~(\ref{eq:M-element})
with the matrix element of spin-flip transitions due to
magneto-dipole interaction, $M_{\rm md} = \mu_B g_e B/2$,
where $g_e$ is the electron $g$-factor and $B$ is magnetic field. 
With $g_e \approx 1.99$ in graphene,  
$M_{\rm md} /  M_{\uparrow \downarrow} \approx 10^{-4}$,
giving %iz which suggests 
a strong SOC enhancement of the spin susceptibility compared to the ESR.  

Our results for the linear spectrum and 
$\alpha_{\rm sf}^{\rm max}(\mu)$ from Eq.~(\ref{eq:absorb}) 
in Fig.~\ref{fig:3}(a) reveal  an enhanced absorption when spin and pseudospin resonances approach each other at $\mu\rightarrow{\Delta}$. 
To analyze $\alpha_{\rm sf}(\omega)$ at  
$\mu \approx \Delta$, one needs to treat SOC non-perturbatively and 
account for spin-pseudospin correlations
responsible for interband spin-flip transitions~\cite{inglot2014optical,kumar2021zero} at combined frequencies ($ \Omega_k \pm \Omega_{\rm ex} $).
We further identify an enhanced spin-light coupling with SOC, as 
$\alpha_{\rm sf}^{\rm max} \propto \lambda_{\rm so}^2/\gamma$ from Eq.~(\ref{eq:absorb}). For 
$\lambda_{\rm so} = 2$~meV and $\hbar/\gamma = 70$~ps, $\alpha_{\rm sf}^{\rm max} > 20 \%$, an order of magnitude larger than $\alpha_0$. 
$\alpha_{\rm sf}^{\rm max} \propto \lambda_{\rm so}^2/\gamma$ is sensitive to the spin relaxation rate~\cite{duckheim2006electric,offidani2018microscopic}, 
which might be suppressed by $\bm{\Delta}$~\cite{wu2010spin} (implying an enhanced EDSR efficiency),  
while also having an inherent anisotropy in graphene-based heterostructures~\cite{raes2016determination,raes2017anisotropy}. 
For massive Dirac electrons and 
different $\bm{E}_{\omega}$ polarizations, 
$\alpha_{\rm sf}^{\rm max}(\mu)$ 
is shown in Fig.~\ref{fig:3}{(b)}~\cite{SM}. 
$\alpha_{\rm sf}^{\rm max}$ 
maximum at {$\mu \approx U/2$} {has} 
the magnitude still larger than ESR. 
For both cases, $\alpha_{\rm sf}^{\rm max}$ 
decreases at $\mu \gg \Delta$ 
due to dynamic suppression of spin-pseudospin coupling, see  
the denominator of $\delta \bm{\tau}_\omega, \delta \bm{s}_\omega$ in Eqs.~(\ref{eq:pseudospinRes}) {and} (\ref{eq:spinres}).

As an alternative to spectroscopic studies, 
we propose the electrical detection of resonant spin generation by THz radiation~\cite{SM}.  This 
is based on interfacial spin-to-charge conversion at the graphene-ferromagnet (F) contact. With the proximity-induced $\Delta$ 
in graphene and the spin-dependent interfacial properties, together with the common $\mu$ and charge transfer, the {THz} absorption 
in graphene leads to a nonequilibrium spin polarization and the generation of an interfacial electromotive force (EMF).  This scheme is an extension of the Johnson-Silsbee spin-charge coupling or spin-voltaic effect~\cite{silsbee1980,johnson1985Interfacial,Zutic2002:PRL,Zutic2004:RMP,fabian2007semiconductor} applied to Dirac materials, where EMF can be detected electrically. To preserve graphene’s Dirac spectrum, in addition to an insulating 
or metallic F with h-BN spacer~\cite{lazic2016effective,xu2018spin}, even a direct contact with a metallic F can be suitable~\cite{asshoff2017magnetoresistance}. 
The enhanced spin-to-charge interconversion at the graphene-F interfaces
enables THz optospintronics and graphene THz detection.

We have revealed the role of coupled spin-pseudospin dynamics for the understanding of THz spin susceptibility
in proximitized Dirac materials. The discovered features are universal
for a wide range of vdW heterostructures:
(i) graphene with proximity-induced Zeeman spin splitting by various magnetic substrates~\cite{dahal2014graphene,cattelan2015nature,liu2015atomic,asshoff2017magnetoresistance,wei2016strong,swartz2012integration,wang2015proximity,zollner2020scattering,tang2020magnetic,zhou2018proximity,farooq2019switchable,ghiasi2021electrical,wu2020large},
(ii) vdW hexagonal crystals with gapped spectrum, such as silicene~\cite{kara2012review}, Bi$(111)$~\cite{reis2017bismuthene},
or puckered 2D lattice with Dirac points~\cite{lu2016multiple},
(iii) nonmagnetic bilayers, such as graphene/TMD~\cite{gmitra2015graphene,garcia2018spin,david2019induced}, with
valley-dependent Zeeman spin splitting due to
the hybridization of graphene $p$-states with TMD bands~\cite{gmitra2015graphene,garcia2018spin,sierra2021van}. 
In the latter case, we predict that for $\mu \lesssim 2U$ and $B=0$, 
the EDSR will be induced selectively for $K$ or $K^\prime$,  depending on the 
circular polarization.
Furthermore, EDSR in graphene/TMD can imprint many-body effects from collective modes of spin-orbital Fermi-liquids~\cite{kumar2021zero,maslov2022collective}.

With challenges and experimental surprises in the understanding of Zeeman
splitting~\cite{Choi2023:NM,Zhou2023:NM}, a key parameter in proximitized vdW heterostructures, 
EDSR studies offer a versatile probe to address this situation and to quantify other proximity-induced spin splittings.
For instance, our predicted polarization structure of $\alpha_{\rm sf}$, with its small-$\mu$ enhancement, has a clear difference as compared to $k$-linear SOC. 
In that case, for the spin-light coupling,
$\bm{k} \to \bm{k} - (e/c) \bm{A}$ in the
$\bm{k}$-linear Rashba Hamiltonian, $H_{\rm R} = \lambda_{\rm R} (\bm{k} \times \bm{\hat}{\bm z} )  \bm{s}$,
leads to the interaction potential, 
$\mathcal{V}' = e\lambda_{\rm R}\left( \bm{\hat}{\bm z}  \times \bm{A}_\omega \right)  \bm{s}/c$,
that couples $\bm{A}_\omega$ directly with the electron spin, rather than with pseudospin. For the usual SOC strength, $\lambda_{\rm R}$, 
the corresponding torque leads to spin-flip transitions in 
$\bm{\Delta}$ with an ordinary polarization structure.
As a fingerprint for different contributions to spin-light coupling
in proximitized Dirac materials, it is natural to analyze the 
polarization dependence of EDSR for different mutual orientations of $\bm{\Delta}$, $\bm{E}_\omega$, and its $\mu$ dependence.

The phenomenon of a coupled spin-pseudospin dynamics has a broad range of implications beyond the EDSR, as it is inherent to many other manifestations of the spin-charge conversion,  such as the spin-voltaic or spin-galvanic effects~\cite{Zutic2002:PRL,Ganichev2002:N}, 
which can be strongly modified in Dirac systems and whose dynamical properties %are not %
remain to be 
understood.
%which are either unexplored in Dirac systems or their dynamical properties remain to be understood. 
Another striking example is the %extensive 
%IZ research on 
study of the spin-orbit torque (SOT), expected to enable the next generation of embedded memories using 2D materials or to integrate photonics, electronics, and 
spintronics~\cite{Yang2022:N,Dainone2024:N,Tsymbal:2019}. %FE
%~\cite{Yang2022:N,Tsymbal:2019,Dainone2024:N}. %FE
%However, the spin-pseudospin coupled SOT in Fig.~2 was previously unknown. 
However, the spin-pseudospin dynamics in %manifestation of 
SOT has not been explored. 
%completely overlooked. 
With the push towards ultrafast SOT~\cite{Jhuria2020:NE}, our analysis of the THz spin-charge conversion, a key SOT ingredient, provides a further 
motivation to consider proximitized vdW-heterostuctures, 
both for the resonant SOT generation and for the THz-spintronics beyond magnetic multilayers.

Our picture could be used to analyze 
nonlinear optical response of Dirac systems and the nonlinear Hall effect~\cite{sodemann2015quantum} for different topological regimes~\cite{malla2021emerging}. 
The inverse effect of spin precession on orbital dynamics can be derived from the coupled spin-pseudospin dynamics, providing an alternative treatment of the topology-sensitive 
Kerr effect~\cite{denisov2022electric,yao2004first}. The discussed picture could be also used to analyze spin-flip transitions in Dirac system with a stronger %IZ larger 
SOC, %IZ  magnitude, 
such as 
graphene-TMD heterostructures~\cite{sierra2021van}, and be implemented in graphene quantum dots and nanoflakes, to realize qubits for THz quantum computing. While SOC 
has been employed to realize fast qubit rotations and control with electric fields~\cite{nadj-perge2010qubit,Wang2022:NC,Stano2022:NRM}, EDSR has not been exploited 
in graphene or bilayer graphene due to their low intrinsic SOC~\cite{PhysRevB.106.245141}. 

We thank 
K. Y. Golenitskii for discussing optical properties of magnetic vdW heterostructures, B. D. McCombe for discussing the EDSR history, and 
D. Torres (ICN2) for help {with} 3D device models in Figs.~1 and 2. Analytical work at Ioffe Institute by K.S.D. 
was supported by the Russian Science Foundation (RSF), project No.22-22-20082 (RSF contract No.22-22-20082 from 25.03.2022, 
contract with a regional grant No.23/2022 from 14.04.2022). I.\v{Z}. was supported by the U.S. DOE, Office of Science BES, Award No. DE-SC0004890. 
S.O.V. was supported by  the Spanish Ministry of Science and Innovation (MCIN) and the Spanish Research Agency (AEI/10.13039/501100011033) through grants 
PID2019-111773RB-I00 and Severo Ochoa CEX2021-001214-S. I.V.R. acknowledges Council for At-Risk Academics (CARA)  support.

\bibliography{Ref2}

\begin{thebibliography}{96}%
\makeatletter
\providecommand \@ifxundefined [1]{%
 \@ifx{#1\undefined}
}%
\providecommand \@ifnum [1]{%
 \ifnum #1\expandafter \@firstoftwo
 \else \expandafter \@secondoftwo
 \fi
}%
\providecommand \@ifx [1]{%
 \ifx #1\expandafter \@firstoftwo
 \else \expandafter \@secondoftwo
 \fi
}%
\providecommand \natexlab [1]{#1}%
\providecommand \enquote  [1]{``#1''}%
\providecommand \bibnamefont  [1]{#1}%
\providecommand \bibfnamefont [1]{#1}%
\providecommand \citenamefont [1]{#1}%
\providecommand \href@noop [0]{\@secondoftwo}%
\providecommand \href [0]{\begingroup \@sanitize@url \@href}%
\providecommand \@href[1]{\@@startlink{#1}\@@href}%
\providecommand \@@href[1]{\endgroup#1\@@endlink}%
\providecommand \@sanitize@url [0]{\catcode `\\12\catcode `\$12\catcode
  `\&12\catcode `\#12\catcode `\^12\catcode `\_12\catcode `\%12\relax}%
\providecommand \@@startlink[1]{}%
\providecommand \@@endlink[0]{}%
\providecommand \url  [0]{\begingroup\@sanitize@url \@url }%
\providecommand \@url [1]{\endgroup\@href {#1}{\urlprefix }}%
\providecommand \urlprefix  [0]{URL }%
\providecommand \Eprint [0]{\href }%
\providecommand \doibase [0]{http://dx.doi.org/}%
\providecommand \selectlanguage [0]{\@gobble}%
\providecommand \bibinfo  [0]{\@secondoftwo}%
\providecommand \bibfield  [0]{\@secondoftwo}%
\providecommand \translation [1]{[#1]}%
\providecommand \BibitemOpen [0]{}%
\providecommand \bibitemStop [0]{}%
\providecommand \bibitemNoStop [0]{.\EOS\space}%
\providecommand \EOS [0]{\spacefactor3000\relax}%
\providecommand \BibitemShut  [1]{\csname bibitem#1\endcsname}%
\let\auto@bib@innerbib\@empty
%</preamble>
\bibitem [{\citenamefont {Geim}\ and\ \citenamefont
  {Grigorieva}(2013)}]{geim2013van}%
  \BibitemOpen
  \bibfield  {author} {\bibinfo {author} {\bibfnamefont {A.~K.}\ \bibnamefont
  {Geim}}\ and\ \bibinfo {author} {\bibfnamefont {I.~V.}\ \bibnamefont
  {Grigorieva}},\ }\bibfield  {title} {\enquote {\bibinfo {title} {van der
  {Waals} heterostructures},}\ }\href {\doibase 10.1038/nature12385} {\bibfield
   {journal} {\bibinfo  {journal} {Nature}\ }\textbf {\bibinfo {volume}
  {499}},\ \bibinfo {pages} {419} (\bibinfo {year} {2013})}\BibitemShut
  {NoStop}%
\bibitem [{\citenamefont {Novoselov}\ \emph {et~al.}(2016)\citenamefont
  {Novoselov}, \citenamefont {Mishchenko}, \citenamefont {Carvalho},\ and\
  \citenamefont {Castro~Neto}}]{novoselov20162d}%
  \BibitemOpen
  \bibfield  {author} {\bibinfo {author} {\bibfnamefont {K.~S.}\ \bibnamefont
  {Novoselov}}, \bibinfo {author} {\bibfnamefont {O.~A.}\ \bibnamefont
  {Mishchenko}}, \bibinfo {author} {\bibfnamefont {O.~A.}\ \bibnamefont
  {Carvalho}}, \ and\ \bibinfo {author} {\bibfnamefont {A.~H.}\ \bibnamefont
  {Castro~Neto}},\ }\bibfield  {title} {\enquote {\bibinfo {title} {{2D}
  materials and van der {Waals} heterostructures},}\ }\href {\doibase
  10.1126/science.aac9439} {\bibfield  {journal} {\bibinfo  {journal}
  {Science}\ }\textbf {\bibinfo {volume} {353}},\ \bibinfo {pages} {461}
  (\bibinfo {year} {2016})}\BibitemShut {NoStop}%
\bibitem [{\citenamefont {{\v{Z}}uti{\'c}}\ \emph {et~al.}(2019)\citenamefont
  {{\v{Z}}uti{\'c}}, \citenamefont {Matos-Abiague}, \citenamefont {Scharf},
  \citenamefont {Dery},\ and\ \citenamefont
  {Belashchenko}}]{vzutic2019proximitized}%
  \BibitemOpen
  \bibfield  {author} {\bibinfo {author} {\bibfnamefont {I.}~\bibnamefont
  {{\v{Z}}uti{\'c}}}, \bibinfo {author} {\bibfnamefont {A.}~\bibnamefont
  {Matos-Abiague}}, \bibinfo {author} {\bibfnamefont {B.}~\bibnamefont
  {Scharf}}, \bibinfo {author} {\bibfnamefont {H.}~\bibnamefont {Dery}}, \ and\
  \bibinfo {author} {\bibfnamefont {K.}~\bibnamefont {Belashchenko}},\
  }\bibfield  {title} {\enquote {\bibinfo {title} {Proximitized materials},}\
  }\href {\doibase 10.1016/j.mattod.2018.05.003} {\bibfield  {journal}
  {\bibinfo  {journal} {Mater. Today}\ }\textbf {\bibinfo {volume} {22}},\
  \bibinfo {pages} {85} (\bibinfo {year} {2019})}\BibitemShut {NoStop}%
\bibitem [{\citenamefont {Sierra}\ \emph {et~al.}(2021)\citenamefont {Sierra},
  \citenamefont {Fabian}, \citenamefont {Kawakami}, \citenamefont {Roche},\
  and\ \citenamefont {Valenzuela}}]{sierra2021van}%
  \BibitemOpen
  \bibfield  {author} {\bibinfo {author} {\bibfnamefont {J.~F.}\ \bibnamefont
  {Sierra}}, \bibinfo {author} {\bibfnamefont {J.}~\bibnamefont {Fabian}},
  \bibinfo {author} {\bibfnamefont {R.~K.}\ \bibnamefont {Kawakami}}, \bibinfo
  {author} {\bibfnamefont {S.}~\bibnamefont {Roche}}, \ and\ \bibinfo {author}
  {\bibfnamefont {S.~O.}\ \bibnamefont {Valenzuela}},\ }\bibfield  {title}
  {\enquote {\bibinfo {title} {{Van} der {Waals} heterostructures for
  spintronics and opto-spintronics},}\ }\href {\doibase
  10.1038/s41565-021-00936-x} {\bibfield  {journal} {\bibinfo  {journal} {Nat.
  Nanotechnol.}\ }\textbf {\bibinfo {volume} {16}},\ \bibinfo {pages} {856}
  (\bibinfo {year} {2021})}\BibitemShut {NoStop}%
\bibitem [{\citenamefont {Han}\ \emph {et~al.}(2014)\citenamefont {Han},
  \citenamefont {Kawakami}, \citenamefont {Gmitra},\ and\ \citenamefont
  {Fabian}}]{han2014graphene}%
  \BibitemOpen
  \bibfield  {author} {\bibinfo {author} {\bibfnamefont {W.}~\bibnamefont
  {Han}}, \bibinfo {author} {\bibfnamefont {R.~K.}\ \bibnamefont {Kawakami}},
  \bibinfo {author} {\bibfnamefont {M.}~\bibnamefont {Gmitra}}, \ and\ \bibinfo
  {author} {\bibfnamefont {J.}~\bibnamefont {Fabian}},\ }\bibfield  {title}
  {\enquote {\bibinfo {title} {Graphene spintronics},}\ }\href {\doibase
  10.1038/nnano.2014.214} {\bibfield  {journal} {\bibinfo  {journal} {Nat.
  Nanotechnol.}\ }\textbf {\bibinfo {volume} {9}},\ \bibinfo {pages} {794}
  (\bibinfo {year} {2014})}\BibitemShut {NoStop}%
\bibitem [{\citenamefont {Han}(2016)}]{han2016perspectives}%
  \BibitemOpen
  \bibfield  {author} {\bibinfo {author} {\bibfnamefont {W.}~\bibnamefont
  {Han}},\ }\bibfield  {title} {\enquote {\bibinfo {title} {Perspectives for
  spintronics in {2D} materials},}\ }\href {\doibase 10.1063/1.4941712}
  {\bibfield  {journal} {\bibinfo  {journal} {APL Mater.}\ }\textbf {\bibinfo
  {volume} {4}},\ \bibinfo {pages} {032401} (\bibinfo {year}
  {2016})}\BibitemShut {NoStop}%
\bibitem [{\citenamefont {Avsar}\ \emph {et~al.}(2020)\citenamefont {Avsar},
  \citenamefont {Ochoa}, \citenamefont {Guinea}, \citenamefont {\"Ozyilmaz},
  \citenamefont {van Wees},\ and\ \citenamefont
  {Vera-Marun}}]{avsar2020colloquium}%
  \BibitemOpen
  \bibfield  {author} {\bibinfo {author} {\bibfnamefont {A.}~\bibnamefont
  {Avsar}}, \bibinfo {author} {\bibfnamefont {H.}~\bibnamefont {Ochoa}},
  \bibinfo {author} {\bibfnamefont {F.}~\bibnamefont {Guinea}}, \bibinfo
  {author} {\bibfnamefont {B.}~\bibnamefont {\"Ozyilmaz}}, \bibinfo {author}
  {\bibfnamefont {B.~J.}\ \bibnamefont {van Wees}}, \ and\ \bibinfo {author}
  {\bibfnamefont {I.~J.}\ \bibnamefont {Vera-Marun}},\ }\bibfield  {title}
  {\enquote {\bibinfo {title} {Colloquium: {Spintronics} in {graphene} and
  {other} {two}-{dimensional} {materials}},}\ }\href {\doibase
  10.1103/RevModPhys.92.021003} {\bibfield  {journal} {\bibinfo  {journal}
  {Rev. Mod. Phys.}\ }\textbf {\bibinfo {volume} {92}},\ \bibinfo {pages}
  {021003} (\bibinfo {year} {2020})}\BibitemShut {NoStop}%
\bibitem [{\citenamefont {Gmitra}\ and\ \citenamefont
  {Fabian}(2015)}]{gmitra2015graphene}%
  \BibitemOpen
  \bibfield  {author} {\bibinfo {author} {\bibfnamefont {M.}~\bibnamefont
  {Gmitra}}\ and\ \bibinfo {author} {\bibfnamefont {J.}~\bibnamefont
  {Fabian}},\ }\bibfield  {title} {\enquote {\bibinfo {title} {Graphene on
  {Transition}-{metal} {dichalcogenides}: A {platform} for {proximity}
  {spin}-{orbit} {physics} and {optospintronics}},}\ }\href {\doibase
  10.1103/PhysRevB.92.155403} {\bibfield  {journal} {\bibinfo  {journal} {Phys.
  Rev. B}\ }\textbf {\bibinfo {volume} {92}},\ \bibinfo {pages} {155403}
  (\bibinfo {year} {2015})}\BibitemShut {NoStop}%
\bibitem [{\citenamefont {Gmitra}\ \emph {et~al.}(2016)\citenamefont {Gmitra},
  \citenamefont {Kochan}, \citenamefont {H\"{o}gl},\ and\ \citenamefont
  {Fabian}}]{gmitra2016graphene}%
  \BibitemOpen
  \bibfield  {author} {\bibinfo {author} {\bibfnamefont {M.}~\bibnamefont
  {Gmitra}}, \bibinfo {author} {\bibfnamefont {D.}~\bibnamefont {Kochan}},
  \bibinfo {author} {\bibfnamefont {P.}~\bibnamefont {H\"{o}gl}}, \ and\
  \bibinfo {author} {\bibfnamefont {J.}~\bibnamefont {Fabian}},\ }\bibfield
  {title} {\enquote {\bibinfo {title} {Trivial and {inverted} {Dirac} {bands}
  and the {emergence} of {quantum} {spin} {Hall} {states} in {graphene} on
  {transition}-{metal} {dichalcogenides}},}\ }\href {\doibase
  10.1103/PhysRevB.93.155104} {\bibfield  {journal} {\bibinfo  {journal} {Phys.
  Rev. B}\ }\textbf {\bibinfo {volume} {93}},\ \bibinfo {pages} {155104}
  (\bibinfo {year} {2016})}\BibitemShut {NoStop}%
\bibitem [{\citenamefont {Garcia}\ \emph {et~al.}(2018)\citenamefont {Garcia},
  \citenamefont {Vila}, \citenamefont {Cummings},\ and\ \citenamefont
  {Roche}}]{garcia2018spin}%
  \BibitemOpen
  \bibfield  {author} {\bibinfo {author} {\bibfnamefont {J.~H.}\ \bibnamefont
  {Garcia}}, \bibinfo {author} {\bibfnamefont {M.}~\bibnamefont {Vila}},
  \bibinfo {author} {\bibfnamefont {A.~W}\ \bibnamefont {Cummings}}, \ and\
  \bibinfo {author} {\bibfnamefont {S.}~\bibnamefont {Roche}},\ }\bibfield
  {title} {\enquote {\bibinfo {title} {Spin transport in graphene/transition
  metal dichalcogenide heterostructures},}\ }\href {\doibase
  10.1039/C7CS00864C} {\bibfield  {journal} {\bibinfo  {journal} {Chem. Soc.
  Rev.}\ }\textbf {\bibinfo {volume} {47}},\ \bibinfo {pages} {3359} (\bibinfo
  {year} {2018})}\BibitemShut {NoStop}%
\bibitem [{\citenamefont {David}\ \emph {et~al.}(2019)\citenamefont {David},
  \citenamefont {Rakyta}, \citenamefont {Korm{\'a}nyos},\ and\ \citenamefont
  {Burkard}}]{david2019induced}%
  \BibitemOpen
  \bibfield  {author} {\bibinfo {author} {\bibfnamefont {A.}~\bibnamefont
  {David}}, \bibinfo {author} {\bibfnamefont {P.}~\bibnamefont {Rakyta}},
  \bibinfo {author} {\bibfnamefont {A.}~\bibnamefont {Korm{\'a}nyos}}, \ and\
  \bibinfo {author} {\bibfnamefont {G.}~\bibnamefont {Burkard}},\ }\bibfield
  {title} {\enquote {\bibinfo {title} {Induced spin-orbit coupling in twisted
  graphene--transition metal dichalcogenide heterobilayers: Twistronics meets
  spintronics},}\ }\href {\doibase 10.1103/PhysRevB.100.085412} {\bibfield
  {journal} {\bibinfo  {journal} {Phys. Rev. B}\ }\textbf {\bibinfo {volume}
  {100}},\ \bibinfo {pages} {085412} (\bibinfo {year} {2019})}\BibitemShut
  {NoStop}%
\bibitem [{\citenamefont {Cummings}\ \emph {et~al.}(2017)\citenamefont
  {Cummings}, \citenamefont {Garcia}, \citenamefont {Fabian},\ and\
  \citenamefont {Roche}}]{cummings2017anisotropy}%
  \BibitemOpen
  \bibfield  {author} {\bibinfo {author} {\bibfnamefont {A.~W.}\ \bibnamefont
  {Cummings}}, \bibinfo {author} {\bibfnamefont {J.~H.}\ \bibnamefont
  {Garcia}}, \bibinfo {author} {\bibfnamefont {J.}~\bibnamefont {Fabian}}, \
  and\ \bibinfo {author} {\bibfnamefont {S.}~\bibnamefont {Roche}},\ }\bibfield
   {title} {\enquote {\bibinfo {title} {Anisotropy in {Graphene} {Induced} by
  {Proximity} {Effects}},}\ }\href {\doibase 10.1103/PhysRevLett.119.206601}
  {\bibfield  {journal} {\bibinfo  {journal} {Phys. Rev. Lett.}\ }\textbf
  {\bibinfo {volume} {119}},\ \bibinfo {pages} {206601} (\bibinfo {year}
  {2017})}\BibitemShut {NoStop}%
\bibitem [{\citenamefont {Ben\'{i}tez}\ \emph {et~al.}(2018)\citenamefont
  {Ben\'{i}tez}, \citenamefont {Sierra}, \citenamefont {Savero~Torres},
  \citenamefont {Arrighi}, \citenamefont {Bonell}, \citenamefont {Costache},\
  and\ \citenamefont {Valenzuela}}]{benitez2018anisotropy}%
  \BibitemOpen
  \bibfield  {author} {\bibinfo {author} {\bibfnamefont {L.~A.}\ \bibnamefont
  {Ben\'{i}tez}}, \bibinfo {author} {\bibfnamefont {J.~F.}\ \bibnamefont
  {Sierra}}, \bibinfo {author} {\bibfnamefont {W.}~\bibnamefont
  {Savero~Torres}}, \bibinfo {author} {\bibfnamefont {A.}~\bibnamefont
  {Arrighi}}, \bibinfo {author} {\bibfnamefont {F.}~\bibnamefont {Bonell}},
  \bibinfo {author} {\bibfnamefont {M.~V.}\ \bibnamefont {Costache}}, \ and\
  \bibinfo {author} {\bibfnamefont {S.~O.}\ \bibnamefont {Valenzuela}},\
  }\bibfield  {title} {\enquote {\bibinfo {title} {Strongly anisotropic spin
  relaxation in graphene–transition metal dichalcogenide heterostructures at
  room temperature},}\ }\href {\doibase 10.1038/s41567-017-0019-2} {\bibfield
  {journal} {\bibinfo  {journal} {Nat. Phys.}\ }\textbf {\bibinfo {volume}
  {14}},\ \bibinfo {pages} {303} (\bibinfo {year} {2018})}\BibitemShut
  {NoStop}%
\bibitem [{\citenamefont {Ghiasi}\ \emph {et~al.}(2017)\citenamefont {Ghiasi},
  \citenamefont {Ingla-Ayn\'{e}s}, \citenamefont {Kaverzin},\ and\
  \citenamefont {van Wees}}]{ghiasi2017anisotropy}%
  \BibitemOpen
  \bibfield  {author} {\bibinfo {author} {\bibfnamefont {T.~S.}\ \bibnamefont
  {Ghiasi}}, \bibinfo {author} {\bibfnamefont {J.}~\bibnamefont
  {Ingla-Ayn\'{e}s}}, \bibinfo {author} {\bibfnamefont {A.~A.}\ \bibnamefont
  {Kaverzin}}, \ and\ \bibinfo {author} {\bibfnamefont {B.~J.}\ \bibnamefont
  {van Wees}},\ }\bibfield  {title} {\enquote {\bibinfo {title} {Induced {Spin}
  {Lifetime} {Anisotropy} in {Transition} {Metal}
  {Dichalcogenide}/{Graphene}},}\ }\href {\doibase
  10.1021/acs.nanolett.7b03460} {\bibfield  {journal} {\bibinfo  {journal}
  {Nano Lett.}\ }\textbf {\bibinfo {volume} {17}},\ \bibinfo {pages} {7528}
  (\bibinfo {year} {2017})}\BibitemShut {NoStop}%
\bibitem [{\citenamefont {Offidani}\ \emph {et~al.}(2017)\citenamefont
  {Offidani}, \citenamefont {Milletar\`{\i}}, \citenamefont {Raimondi},\ and\
  \citenamefont {Ferreira}}]{offidani2017optimal}%
  \BibitemOpen
  \bibfield  {author} {\bibinfo {author} {\bibfnamefont {M.}~\bibnamefont
  {Offidani}}, \bibinfo {author} {\bibfnamefont {M.}~\bibnamefont
  {Milletar\`{\i}}}, \bibinfo {author} {\bibfnamefont {R.}~\bibnamefont
  {Raimondi}}, \ and\ \bibinfo {author} {\bibfnamefont {A.}~\bibnamefont
  {Ferreira}},\ }\bibfield  {title} {\enquote {\bibinfo {title} {Optimal
  {Charge}-to-{Spin} {Conversion} in {Graphene} on {Transition}-{Metal}
  {Dichalcogenides}},}\ }\href {\doibase 10.1103/PhysRevLett.119.196801}
  {\bibfield  {journal} {\bibinfo  {journal} {Phys. Rev. Lett.}\ }\textbf
  {\bibinfo {volume} {119}},\ \bibinfo {pages} {196801} (\bibinfo {year}
  {2017})}\BibitemShut {NoStop}%
\bibitem [{\citenamefont {Garcia}\ \emph {et~al.}(2017)\citenamefont {Garcia},
  \citenamefont {Cummings},\ and\ \citenamefont {Roche}}]{garcia2017SHE}%
  \BibitemOpen
  \bibfield  {author} {\bibinfo {author} {\bibfnamefont {J.~H.}\ \bibnamefont
  {Garcia}}, \bibinfo {author} {\bibfnamefont {A.~W.}\ \bibnamefont
  {Cummings}}, \ and\ \bibinfo {author} {\bibfnamefont {S.}~\bibnamefont
  {Roche}},\ }\bibfield  {title} {\enquote {\bibinfo {title} {{Spin} {Hall}
  {Effect} and {Weak} {Antilocalization} in {Graphene}/{Transition} {Metal}
  {Dichalcogenide} {Heterostructures}},}\ }\href {\doibase
  10.1021/acs.nanolett.7b02364} {\bibfield  {journal} {\bibinfo  {journal}
  {Nano Lett.}\ }\textbf {\bibinfo {volume} {17}},\ \bibinfo {pages} {5078}
  (\bibinfo {year} {2017})}\BibitemShut {NoStop}%
\bibitem [{\citenamefont {Safeer}\ \emph {et~al.}(2019)\citenamefont {Safeer},
  \citenamefont {Ingla-Ayn{\'e}s}, \citenamefont {Herling}, \citenamefont
  {Garcia}, \citenamefont {Vila}, \citenamefont {Ontoso}, \citenamefont
  {Calvo}, \citenamefont {Roche}, \citenamefont {Hueso},\ and\ \citenamefont
  {Casanova}}]{safeer2019MoS2}%
  \BibitemOpen
  \bibfield  {author} {\bibinfo {author} {\bibfnamefont {C.~K.}\ \bibnamefont
  {Safeer}}, \bibinfo {author} {\bibfnamefont {J.}~\bibnamefont
  {Ingla-Ayn{\'e}s}}, \bibinfo {author} {\bibfnamefont {F.}~\bibnamefont
  {Herling}}, \bibinfo {author} {\bibfnamefont {J.~H}\ \bibnamefont {Garcia}},
  \bibinfo {author} {\bibfnamefont {M.}~\bibnamefont {Vila}}, \bibinfo {author}
  {\bibfnamefont {N.}~\bibnamefont {Ontoso}}, \bibinfo {author} {\bibfnamefont
  {M.~R.}\ \bibnamefont {Calvo}}, \bibinfo {author} {\bibfnamefont
  {S.}~\bibnamefont {Roche}}, \bibinfo {author} {\bibfnamefont {L.~E.}\
  \bibnamefont {Hueso}}, \ and\ \bibinfo {author} {\bibfnamefont
  {F.}~\bibnamefont {Casanova}},\ }\bibfield  {title} {\enquote {\bibinfo
  {title} {Room-{Temperature} {Spin} {Hall} {Effect} in {Graphene}/{MoS$_2$}
  van der {Waals} {Heterostructures}},}\ }\href {\doibase
  10.1021/acs.nanolett.8b04368} {\bibfield  {journal} {\bibinfo  {journal}
  {Nano Lett.}\ }\textbf {\bibinfo {volume} {19}},\ \bibinfo {pages} {1074}
  (\bibinfo {year} {2019})}\BibitemShut {NoStop}%
\bibitem [{\citenamefont {Ben\'{i}tez}\ \emph {et~al.}(2020)\citenamefont
  {Ben\'{i}tez}, \citenamefont {Savero~Torres}, \citenamefont {Sierra},
  \citenamefont {Timmermans}, \citenamefont {Garcia}, \citenamefont {Roche},
  \citenamefont {Costache},\ and\ \citenamefont
  {Valenzuela}}]{benitez2020SHEISGE}%
  \BibitemOpen
  \bibfield  {author} {\bibinfo {author} {\bibfnamefont {L.~A.}\ \bibnamefont
  {Ben\'{i}tez}}, \bibinfo {author} {\bibfnamefont {W.}~\bibnamefont
  {Savero~Torres}}, \bibinfo {author} {\bibfnamefont {J.~F.}\ \bibnamefont
  {Sierra}}, \bibinfo {author} {\bibfnamefont {M.}~\bibnamefont {Timmermans}},
  \bibinfo {author} {\bibfnamefont {J.~H.}\ \bibnamefont {Garcia}}, \bibinfo
  {author} {\bibfnamefont {S.}~\bibnamefont {Roche}}, \bibinfo {author}
  {\bibfnamefont {M.~V.}\ \bibnamefont {Costache}}, \ and\ \bibinfo {author}
  {\bibfnamefont {S.~O.}\ \bibnamefont {Valenzuela}},\ }\bibfield  {title}
  {\enquote {\bibinfo {title} {Tunable room-temperature spin galvanic and spin
  {Hall} effects in van der {Waals} heterostructures},}\ }\href {\doibase
  10.1038/s41563-019-0575-1} {\bibfield  {journal} {\bibinfo  {journal} {Nat.
  Mater.}\ }\textbf {\bibinfo {volume} {19}},\ \bibinfo {pages} {107} (\bibinfo
  {year} {2020})}\BibitemShut {NoStop}%
\bibitem [{\citenamefont {Galceran}\ \emph {et~al.}(2021)\citenamefont
  {Galceran}, \citenamefont {Tian}, \citenamefont {Li}, \citenamefont {Bonell},
  \citenamefont {Jamet}, \citenamefont {Vergnaud}, \citenamefont {Marty},
  \citenamefont {Garc{\'\i}a}, \citenamefont {Sierra}, \citenamefont
  {Costache},\ and\ \citenamefont {et~al.}}]{galceran2021CSI}%
  \BibitemOpen
  \bibfield  {author} {\bibinfo {author} {\bibfnamefont {R.}~\bibnamefont
  {Galceran}}, \bibinfo {author} {\bibfnamefont {B.}~\bibnamefont {Tian}},
  \bibinfo {author} {\bibfnamefont {J.}~\bibnamefont {Li}}, \bibinfo {author}
  {\bibfnamefont {F.}~\bibnamefont {Bonell}}, \bibinfo {author} {\bibfnamefont
  {M.}~\bibnamefont {Jamet}}, \bibinfo {author} {\bibfnamefont
  {C.}~\bibnamefont {Vergnaud}}, \bibinfo {author} {\bibfnamefont
  {A.}~\bibnamefont {Marty}}, \bibinfo {author} {\bibfnamefont {J.~H.}\
  \bibnamefont {Garc{\'\i}a}}, \bibinfo {author} {\bibfnamefont {J.~F.}\
  \bibnamefont {Sierra}}, \bibinfo {author} {\bibfnamefont {M.~V.}\
  \bibnamefont {Costache}}, \ and\ \bibinfo {author} {\bibnamefont {et~al.}},\
  }\bibfield  {title} {\enquote {\bibinfo {title} {Control of spin–charge
  conversion in van der {Waals} heterostructures},}\ }\href {\doibase
  10.1063/5.0054865} {\bibfield  {journal} {\bibinfo  {journal} {APL Mater.}\
  }\textbf {\bibinfo {volume} {9}},\ \bibinfo {pages} {100901} (\bibinfo {year}
  {2021})}\BibitemShut {NoStop}%
\bibitem [{\citenamefont {Yang}\ \emph {et~al.}(2013)\citenamefont {Yang},
  \citenamefont {Hallal}, \citenamefont {Terrade}, \citenamefont {Waintal},
  \citenamefont {Roche},\ and\ \citenamefont {Chshiev}}]{yang2013proximity}%
  \BibitemOpen
  \bibfield  {author} {\bibinfo {author} {\bibfnamefont {H.~X.}\ \bibnamefont
  {Yang}}, \bibinfo {author} {\bibfnamefont {A.}~\bibnamefont {Hallal}},
  \bibinfo {author} {\bibfnamefont {D.}~\bibnamefont {Terrade}}, \bibinfo
  {author} {\bibfnamefont {X.}~\bibnamefont {Waintal}}, \bibinfo {author}
  {\bibfnamefont {S.}~\bibnamefont {Roche}}, \ and\ \bibinfo {author}
  {\bibfnamefont {M.}~\bibnamefont {Chshiev}},\ }\bibfield  {title} {\enquote
  {\bibinfo {title} {Proximity {Effects} {Induced} in {Graphene} by {Magnetic}
  {Insulators}: {First}-{Principles} {Calculations} on {Spin} {Filtering} and
  {Exchange}-{Splitting} {Gaps}},}\ }\href {\doibase
  10.1103/PhysRevLett.110.046603} {\bibfield  {journal} {\bibinfo  {journal}
  {Phys. Rev. Lett.}\ }\textbf {\bibinfo {volume} {110}},\ \bibinfo {pages}
  {046603} (\bibinfo {year} {2013})}\BibitemShut {NoStop}%
\bibitem [{\citenamefont {Lazi{\'c}}\ \emph {et~al.}(2016)\citenamefont
  {Lazi{\'c}}, \citenamefont {Belashchenko},\ and\ \citenamefont
  {{\v{Z}}uti{\'c}}}]{lazic2016effective}%
  \BibitemOpen
  \bibfield  {author} {\bibinfo {author} {\bibfnamefont {P.}~\bibnamefont
  {Lazi{\'c}}}, \bibinfo {author} {\bibfnamefont {K.~D.}\ \bibnamefont
  {Belashchenko}}, \ and\ \bibinfo {author} {\bibfnamefont {I.}~\bibnamefont
  {{\v{Z}}uti{\'c}}},\ }\bibfield  {title} {\enquote {\bibinfo {title}
  {Effective {gating} and {tunable} {magnetic} {proximity} {effects} in
  {two}-{dimensional} {heterostructures}},}\ }\href {\doibase
  10.1103/PhysRevB.93.241401} {\bibfield  {journal} {\bibinfo  {journal} {Phys.
  Rev. B}\ }\textbf {\bibinfo {volume} {93}},\ \bibinfo {pages} {241401}
  (\bibinfo {year} {2016})}\BibitemShut {NoStop}%
\bibitem [{\citenamefont {Asshoff}\ \emph {et~al.}(2017)\citenamefont
  {Asshoff}, \citenamefont {Sambricio}, \citenamefont {Rooney}, \citenamefont
  {Slizovskiy}, \citenamefont {Mishchenko}, \citenamefont {Rakowski},
  \citenamefont {Hill}, \citenamefont {Geim}, \citenamefont {Haigh},
  \citenamefont {Fal'ko},\ and\ \citenamefont
  {et~al.}}]{asshoff2017magnetoresistance}%
  \BibitemOpen
  \bibfield  {author} {\bibinfo {author} {\bibfnamefont {P.~U.}\ \bibnamefont
  {Asshoff}}, \bibinfo {author} {\bibfnamefont {J.~L.}\ \bibnamefont
  {Sambricio}}, \bibinfo {author} {\bibfnamefont {A.~P.}\ \bibnamefont
  {Rooney}}, \bibinfo {author} {\bibfnamefont {S.}~\bibnamefont {Slizovskiy}},
  \bibinfo {author} {\bibfnamefont {A.}~\bibnamefont {Mishchenko}}, \bibinfo
  {author} {\bibfnamefont {A.~M.}\ \bibnamefont {Rakowski}}, \bibinfo {author}
  {\bibfnamefont {E.~W.}\ \bibnamefont {Hill}}, \bibinfo {author}
  {\bibfnamefont {A.~K.}\ \bibnamefont {Geim}}, \bibinfo {author}
  {\bibfnamefont {S.~J.}\ \bibnamefont {Haigh}}, \bibinfo {author}
  {\bibfnamefont {V.~I.}\ \bibnamefont {Fal'ko}}, \ and\ \bibinfo {author}
  {\bibnamefont {et~al.}},\ }\bibfield  {title} {\enquote {\bibinfo {title}
  {Magnetoresistance of vertical {Co}-graphene-{NiFe} junctions controlled by
  charge transfer and proximity-induced spin splitting in graphene},}\ }\href
  {\doibase 10.1088/2053-1583/aa7452} {\bibfield  {journal} {\bibinfo
  {journal} {2D Mater.}\ }\textbf {\bibinfo {volume} {4}},\ \bibinfo {pages}
  {031004} (\bibinfo {year} {2017})}\BibitemShut {NoStop}%
\bibitem [{\citenamefont {Wei}\ \emph {et~al.}(2016)\citenamefont {Wei},
  \citenamefont {Lee}, \citenamefont {Lemaitre}, \citenamefont {Pinel},
  \citenamefont {Cutaia}, \citenamefont {Cha}, \citenamefont {Katmis},
  \citenamefont {Zhu}, \citenamefont {Heiman}, \citenamefont {Hone},\ and\
  \citenamefont {et~al.}}]{wei2016strong}%
  \BibitemOpen
  \bibfield  {author} {\bibinfo {author} {\bibfnamefont {P.}~\bibnamefont
  {Wei}}, \bibinfo {author} {\bibfnamefont {S.}~\bibnamefont {Lee}}, \bibinfo
  {author} {\bibfnamefont {F.}~\bibnamefont {Lemaitre}}, \bibinfo {author}
  {\bibfnamefont {L.}~\bibnamefont {Pinel}}, \bibinfo {author} {\bibfnamefont
  {D.}~\bibnamefont {Cutaia}}, \bibinfo {author} {\bibfnamefont
  {W.}~\bibnamefont {Cha}}, \bibinfo {author} {\bibfnamefont {F.}~\bibnamefont
  {Katmis}}, \bibinfo {author} {\bibfnamefont {Y.}~\bibnamefont {Zhu}},
  \bibinfo {author} {\bibfnamefont {D.}~\bibnamefont {Heiman}}, \bibinfo
  {author} {\bibfnamefont {J.}~\bibnamefont {Hone}}, \ and\ \bibinfo {author}
  {\bibnamefont {et~al.}},\ }\bibfield  {title} {\enquote {\bibinfo {title}
  {Strong interfacial exchange field in the graphene/{EuS} heterostructure},}\
  }\href {\doibase 10.1038/nmat4603} {\bibfield  {journal} {\bibinfo  {journal}
  {Nat. Mater.}\ }\textbf {\bibinfo {volume} {15}},\ \bibinfo {pages} {711}
  (\bibinfo {year} {2016})}\BibitemShut {NoStop}%
\bibitem [{\citenamefont {Wu}\ \emph {et~al.}(2020)\citenamefont {Wu},
  \citenamefont {Yin}, \citenamefont {Pan}, \citenamefont {Grutter},
  \citenamefont {Pan}, \citenamefont {Lee}, \citenamefont {Gilbert},
  \citenamefont {Borchers}, \citenamefont {Ratcliff}, \citenamefont {Li},\ and\
  \citenamefont {et~al.}}]{wu2020large}%
  \BibitemOpen
  \bibfield  {author} {\bibinfo {author} {\bibfnamefont {Y.}~\bibnamefont
  {Wu}}, \bibinfo {author} {\bibfnamefont {G.}~\bibnamefont {Yin}}, \bibinfo
  {author} {\bibfnamefont {L.}~\bibnamefont {Pan}}, \bibinfo {author}
  {\bibfnamefont {A.~J.}\ \bibnamefont {Grutter}}, \bibinfo {author}
  {\bibfnamefont {Q.}~\bibnamefont {Pan}}, \bibinfo {author} {\bibfnamefont
  {A.}~\bibnamefont {Lee}}, \bibinfo {author} {\bibfnamefont {D.~A.}\
  \bibnamefont {Gilbert}}, \bibinfo {author} {\bibfnamefont {J.~A.}\
  \bibnamefont {Borchers}}, \bibinfo {author} {\bibfnamefont {W.}~\bibnamefont
  {Ratcliff}}, \bibinfo {author} {\bibfnamefont {A.}~\bibnamefont {Li}}, \ and\
  \bibinfo {author} {\bibnamefont {et~al.}},\ }\bibfield  {title} {\enquote
  {\bibinfo {title} {Large exchange splitting in monolayer graphene magnetized
  by an antiferromagnet},}\ }\href {\doibase 10.1038/s41928-020-0458-0}
  {\bibfield  {journal} {\bibinfo  {journal} {Nat. Electron.}\ }\textbf
  {\bibinfo {volume} {3}},\ \bibinfo {pages} {604} (\bibinfo {year}
  {2020})}\BibitemShut {NoStop}%
\bibitem [{\citenamefont {Ghiasi}\ \emph {et~al.}(2021)\citenamefont {Ghiasi},
  \citenamefont {Kaverzin}, \citenamefont {Dismukes}, \citenamefont {de~Wal},
  \citenamefont {Roy},\ and\ \citenamefont {van Wees}}]{ghiasi2021electrical}%
  \BibitemOpen
  \bibfield  {author} {\bibinfo {author} {\bibfnamefont {T.~S.}\ \bibnamefont
  {Ghiasi}}, \bibinfo {author} {\bibfnamefont {A.~A.}\ \bibnamefont
  {Kaverzin}}, \bibinfo {author} {\bibfnamefont {A.~H.}\ \bibnamefont
  {Dismukes}}, \bibinfo {author} {\bibfnamefont {D.~K.}\ \bibnamefont
  {de~Wal}}, \bibinfo {author} {\bibfnamefont {X.}~\bibnamefont {Roy}}, \ and\
  \bibinfo {author} {\bibfnamefont {B.~J.}\ \bibnamefont {van Wees}},\
  }\bibfield  {title} {\enquote {\bibinfo {title} {Electrical and thermal
  generation of spin currents by magnetic bilayer graphene},}\ }\href {\doibase
  10.1038/s41565-021-00887-3} {\bibfield  {journal} {\bibinfo  {journal} {Nat.
  Nanotechnol.}\ }\textbf {\bibinfo {volume} {16}},\ \bibinfo {pages} {788}
  (\bibinfo {year} {2021})}\BibitemShut {NoStop}%
\bibitem [{\citenamefont {Wen}\ \emph {et~al.}(2016)\citenamefont {Wen},
  \citenamefont {Dery}, \citenamefont {Amamou}, \citenamefont {Zhu},
  \citenamefont {Lin}, \citenamefont {Shi}, \citenamefont {{\v{Z}uti\'c}},
  \citenamefont {Krivorotov}, \citenamefont {Sham},\ and\ \citenamefont
  {Kawakami}}]{Wen2016:PRA}%
  \BibitemOpen
  \bibfield  {author} {\bibinfo {author} {\bibfnamefont {H.}~\bibnamefont
  {Wen}}, \bibinfo {author} {\bibfnamefont {H.}~\bibnamefont {Dery}}, \bibinfo
  {author} {\bibfnamefont {W.}~\bibnamefont {Amamou}}, \bibinfo {author}
  {\bibfnamefont {T.}~\bibnamefont {Zhu}}, \bibinfo {author} {\bibfnamefont
  {Z.}~\bibnamefont {Lin}}, \bibinfo {author} {\bibfnamefont {J.}~\bibnamefont
  {Shi}}, \bibinfo {author} {\bibfnamefont {I.}~\bibnamefont {{\v{Z}uti\'c}}},
  \bibinfo {author} {\bibfnamefont {I.}~\bibnamefont {Krivorotov}}, \bibinfo
  {author} {\bibfnamefont {Lu~J.}\ \bibnamefont {Sham}}, \ and\ \bibinfo
  {author} {\bibfnamefont {R.~K.}\ \bibnamefont {Kawakami}},\ }\bibfield
  {title} {\enquote {\bibinfo {title} {Experimental {Demonstration} of {XOR}
  {Operation} in {Graphene} {Magnetologic} {Gates} at {Room} {Temperature}},}\
  }\href {\doibase {10.1103/PhysRevApplied.5.044003}} {\bibfield  {journal}
  {\bibinfo  {journal} {Phys. Rev. Appl.}\ }\textbf {\bibinfo {volume} {5}},\
  \bibinfo {pages} {{044003}} (\bibinfo {year} {2016})}\BibitemShut {NoStop}%
\bibitem [{\citenamefont {Castilla}\ \emph {et~al.}(2019)\citenamefont
  {Castilla}, \citenamefont {Terr{\'e}s}, \citenamefont {Autore}, \citenamefont
  {Viti}, \citenamefont {Li}, \citenamefont {Nikitin}, \citenamefont
  {Vangelidis}, \citenamefont {Watanabe}, \citenamefont {Taniguchi},
  \citenamefont {Lidorikis},\ and\ \citenamefont {et~al.}}]{castilla2019fast}%
  \BibitemOpen
  \bibfield  {author} {\bibinfo {author} {\bibfnamefont {S.}~\bibnamefont
  {Castilla}}, \bibinfo {author} {\bibfnamefont {B.}~\bibnamefont
  {Terr{\'e}s}}, \bibinfo {author} {\bibfnamefont {M.}~\bibnamefont {Autore}},
  \bibinfo {author} {\bibfnamefont {L.}~\bibnamefont {Viti}}, \bibinfo {author}
  {\bibfnamefont {J.}~\bibnamefont {Li}}, \bibinfo {author} {\bibfnamefont
  {A.~Y.}\ \bibnamefont {Nikitin}}, \bibinfo {author} {\bibfnamefont
  {I.}~\bibnamefont {Vangelidis}}, \bibinfo {author} {\bibfnamefont
  {K.}~\bibnamefont {Watanabe}}, \bibinfo {author} {\bibfnamefont
  {T.}~\bibnamefont {Taniguchi}}, \bibinfo {author} {\bibfnamefont
  {E.}~\bibnamefont {Lidorikis}}, \ and\ \bibinfo {author} {\bibnamefont
  {et~al.}},\ }\bibfield  {title} {\enquote {\bibinfo {title} {Fast and
  {Sensitive} {Terahertz} {Detection} {Using} an {Antenna}-{Integrated}
  {Graphene} pn {Junction}},}\ }\href {\doibase 10.1021/acs.nanolett.8b04171}
  {\bibfield  {journal} {\bibinfo  {journal} {Nano Lett.}\ }\textbf {\bibinfo
  {volume} {19}},\ \bibinfo {pages} {2765} (\bibinfo {year}
  {2019})}\BibitemShut {NoStop}%
\bibitem [{\citenamefont {Hafez}\ \emph {et~al.}(2020)\citenamefont {Hafez},
  \citenamefont {Kovalev}, \citenamefont {Tielrooij}, \citenamefont {Bonn},
  \citenamefont {Gensch},\ and\ \citenamefont {Turchinovich}}]{hafez2020THz}%
  \BibitemOpen
  \bibfield  {author} {\bibinfo {author} {\bibfnamefont {H.~A.}\ \bibnamefont
  {Hafez}}, \bibinfo {author} {\bibfnamefont {S.}~\bibnamefont {Kovalev}},
  \bibinfo {author} {\bibfnamefont {K.}~\bibnamefont {Tielrooij}}, \bibinfo
  {author} {\bibfnamefont {M.}~\bibnamefont {Bonn}}, \bibinfo {author}
  {\bibfnamefont {M.}~\bibnamefont {Gensch}}, \ and\ \bibinfo {author}
  {\bibfnamefont {D.}~\bibnamefont {Turchinovich}},\ }\bibfield  {title}
  {\enquote {\bibinfo {title} {Terahertz {Nonlinear} {Optics} of {Graphene}:
  {From} {Saturable} {Absorption} to {High}-{Harmonics} {Generation}},}\ }\href
  {\doibase 10.1002/adom.201900771} {\bibfield  {journal} {\bibinfo  {journal}
  {Adv. Opt. Mater.}\ }\textbf {\bibinfo {volume} {8}},\ \bibinfo {pages}
  {1900771} (\bibinfo {year} {2020})}\BibitemShut {NoStop}%
\bibitem [{\citenamefont {Kovalev}\ \emph {et~al.}(2021)\citenamefont
  {Kovalev}, \citenamefont {Hafez}, \citenamefont {Tielrooij}, \citenamefont
  {Deinert}, \citenamefont {Ilyakov}, \citenamefont {Awari}, \citenamefont
  {Alcaraz}, \citenamefont {Soundarapandian}, \citenamefont {Saleta},
  \citenamefont {Germanskiy},\ and\ \citenamefont {et~al.}}]{kovalev2021THz}%
  \BibitemOpen
  \bibfield  {author} {\bibinfo {author} {\bibfnamefont {S.}~\bibnamefont
  {Kovalev}}, \bibinfo {author} {\bibfnamefont {H.~A.}\ \bibnamefont {Hafez}},
  \bibinfo {author} {\bibfnamefont {K.}~\bibnamefont {Tielrooij}}, \bibinfo
  {author} {\bibfnamefont {J.}~\bibnamefont {Deinert}}, \bibinfo {author}
  {\bibfnamefont {I.}~\bibnamefont {Ilyakov}}, \bibinfo {author} {\bibfnamefont
  {N.}~\bibnamefont {Awari}}, \bibinfo {author} {\bibfnamefont
  {D.}~\bibnamefont {Alcaraz}}, \bibinfo {author} {\bibfnamefont
  {K.}~\bibnamefont {Soundarapandian}}, \bibinfo {author} {\bibfnamefont
  {D.}~\bibnamefont {Saleta}}, \bibinfo {author} {\bibfnamefont
  {S.}~\bibnamefont {Germanskiy}}, \ and\ \bibinfo {author} {\bibnamefont
  {et~al.}},\ }\bibfield  {title} {\enquote {\bibinfo {title} {Electrical
  tunability of terahertz nonlinearity in graphene},}\ }\href {\doibase
  10.1126/sciadv.abf9809} {\bibfield  {journal} {\bibinfo  {journal} {Sci.
  Adv.}\ }\textbf {\bibinfo {volume} {7}},\ \bibinfo {pages} {eabf9809}
  (\bibinfo {year} {2021})}\BibitemShut {NoStop}%
\bibitem [{\citenamefont {Tielrooij}\ \emph {et~al.}(2022)\citenamefont
  {Tielrooij}, \citenamefont {Principi}, \citenamefont {Reig}, \citenamefont
  {Block}, \citenamefont {Varghese}, \citenamefont {Schreyeck}, \citenamefont
  {Brunner}, \citenamefont {Karczewski}, \citenamefont {Ilyakov}, \citenamefont
  {Ponomaryov},\ and\ \citenamefont {et~al.}}]{tielrooij2021THz}%
  \BibitemOpen
  \bibfield  {author} {\bibinfo {author} {\bibfnamefont {K.-J.}\ \bibnamefont
  {Tielrooij}}, \bibinfo {author} {\bibfnamefont {A.}~\bibnamefont {Principi}},
  \bibinfo {author} {\bibfnamefont {D.~S.}\ \bibnamefont {Reig}}, \bibinfo
  {author} {\bibfnamefont {A.}~\bibnamefont {Block}}, \bibinfo {author}
  {\bibfnamefont {S.}~\bibnamefont {Varghese}}, \bibinfo {author}
  {\bibfnamefont {S.}~\bibnamefont {Schreyeck}}, \bibinfo {author}
  {\bibfnamefont {K.}~\bibnamefont {Brunner}}, \bibinfo {author} {\bibfnamefont
  {G.}~\bibnamefont {Karczewski}}, \bibinfo {author} {\bibfnamefont
  {I.}~\bibnamefont {Ilyakov}}, \bibinfo {author} {\bibfnamefont
  {O.}~\bibnamefont {Ponomaryov}}, \ and\ \bibinfo {author} {\bibnamefont
  {et~al.}},\ }\bibfield  {title} {\enquote {\bibinfo {title} {Milliwatt
  terahertz harmonic generation from topological insulator metamaterials},}\
  }\href {\doibase 10.1038/s41377-022-01008-y} {\bibfield  {journal} {\bibinfo
  {journal} {Light Sci. Appl}\ }\textbf {\bibinfo {volume} {11}},\ \bibinfo
  {pages} {315} (\bibinfo {year} {2022})}\BibitemShut {NoStop}%
\bibitem [{\citenamefont {Rashba}(1960)}]{rashba1960properties}%
  \BibitemOpen
  \bibfield  {author} {\bibinfo {author} {\bibfnamefont {E.~I.}\ \bibnamefont
  {Rashba}},\ }\bibfield  {title} {\enquote {\bibinfo {title} {Properties of
  semiconductors with an extremum loop. $\mathrm{I}$. {Cyclotron} and
  combinational resonance in a magnetic field perpendicular to the plane of the
  loop},}\ }\href@noop {} {\bibfield  {journal} {\bibinfo  {journal} {Sov.
  Phys. Solid State}\ }\textbf {\bibinfo {volume} {2}},\ \bibinfo {pages}
  {1109} (\bibinfo {year} {1960})}\BibitemShut {NoStop}%
\bibitem [{\citenamefont {Rashba}(2002)}]{Rashba2002:JS}%
  \BibitemOpen
  \bibfield  {author} {\bibinfo {author} {\bibfnamefont {E.~I.}\ \bibnamefont
  {Rashba}},\ }\bibfield  {title} {\enquote {\bibinfo {title} {Spintronics:
  {Sources} and {Challenge}},}\ }\href {\doibase 10.1023/A:1014066808432}
  {\bibfield  {journal} {\bibinfo  {journal} {J. Supercond.}\ }\textbf
  {\bibinfo {volume} {15}},\ \bibinfo {pages} {13} (\bibinfo {year}
  {2002})}\BibitemShut {NoStop}%
\bibitem [{\citenamefont {Bell}(1962)}]{Bell1962:PRL}%
  \BibitemOpen
  \bibfield  {author} {\bibinfo {author} {\bibfnamefont {R.~L.}\ \bibnamefont
  {Bell}},\ }\bibfield  {title} {\enquote {\bibinfo {title} {Electric {Dipole}
  {Spin} {Transitions} in {InSb}},}\ }\href {\doibase 10.1103/PhysRevLett.9.52}
  {\bibfield  {journal} {\bibinfo  {journal} {Phys. Rev. Lett.}\ }\textbf
  {\bibinfo {volume} {9}},\ \bibinfo {pages} {52} (\bibinfo {year}
  {1962})}\BibitemShut {NoStop}%
\bibitem [{\citenamefont {McCombe}\ \emph {et~al.}(1967)\citenamefont
  {McCombe}, \citenamefont {Bishop},\ and\ \citenamefont
  {Kaplan}}]{McCombe1967:PRL}%
  \BibitemOpen
  \bibfield  {author} {\bibinfo {author} {\bibfnamefont {B.~D.}\ \bibnamefont
  {McCombe}}, \bibinfo {author} {\bibfnamefont {S.~G.}\ \bibnamefont {Bishop}},
  \ and\ \bibinfo {author} {\bibfnamefont {R.}~\bibnamefont {Kaplan}},\
  }\bibfield  {title} {\enquote {\bibinfo {title} {Combined {Resonance} and
  {Electron} $g$ {Values} in {InSb}},}\ }\href {\doibase
  10.1103/PhysRevLett.18.748} {\bibfield  {journal} {\bibinfo  {journal} {Phys.
  Rev. Lett.}\ }\textbf {\bibinfo {volume} {18}},\ \bibinfo {pages} {748}
  (\bibinfo {year} {1967})}\BibitemShut {NoStop}%
\bibitem [{\citenamefont {Pekar}\ and\ \citenamefont
  {Rashba}(1965)}]{Pekar1964:SPJETP}%
  \BibitemOpen
  \bibfield  {author} {\bibinfo {author} {\bibfnamefont {S.~I.}\ \bibnamefont
  {Pekar}}\ and\ \bibinfo {author} {\bibfnamefont {E.~I.}\ \bibnamefont
  {Rashba}},\ }\bibfield  {title} {\enquote {\bibinfo {title} {Combined
  resonance in crystals in inhomogeneous magnetic field},}\ }\href@noop {}
  {\bibfield  {journal} {\bibinfo  {journal} {Sov. Phys. JETP}\ }\textbf
  {\bibinfo {volume} {20}},\ \bibinfo {pages} {1295} (\bibinfo {year}
  {1965})}\BibitemShut {NoStop}%
\bibitem [{\citenamefont {{\v{Z}uti\'{c}}}\ \emph {et~al.}(2004)\citenamefont
  {{\v{Z}uti\'{c}}}, \citenamefont {Fabian},\ and\ \citenamefont {{Das
  Sarma}}}]{Zutic2004:RMP}%
  \BibitemOpen
  \bibfield  {author} {\bibinfo {author} {\bibfnamefont {I.}~\bibnamefont
  {{\v{Z}uti\'{c}}}}, \bibinfo {author} {\bibfnamefont {J.}~\bibnamefont
  {Fabian}}, \ and\ \bibinfo {author} {\bibfnamefont {S.}~\bibnamefont {{Das
  Sarma}}},\ }\bibfield  {title} {\enquote {\bibinfo {title} {Spintronics:
  {Fundamentals} and {applications}},}\ }\href {\doibase
  10.1103/RevModPhys.76.323} {\bibfield  {journal} {\bibinfo  {journal} {Rev.
  Mod. Phys.}\ }\textbf {\bibinfo {volume} {76}},\ \bibinfo {pages} {323}
  (\bibinfo {year} {2004})}\BibitemShut {NoStop}%
\bibitem [{\citenamefont {Duckheim}\ and\ \citenamefont
  {Loss}(2006)}]{duckheim2006electric}%
  \BibitemOpen
  \bibfield  {author} {\bibinfo {author} {\bibfnamefont {M.}~\bibnamefont
  {Duckheim}}\ and\ \bibinfo {author} {\bibfnamefont {D.}~\bibnamefont
  {Loss}},\ }\bibfield  {title} {\enquote {\bibinfo {title}
  {Electric-dipole-induced spin resonance in disordered semiconductors},}\
  }\href {\doibase 10.1038/nphys238} {\bibfield  {journal} {\bibinfo  {journal}
  {Nat. Phys.}\ }\textbf {\bibinfo {volume} {2}},\ \bibinfo {pages} {195}
  (\bibinfo {year} {2006})}\BibitemShut {NoStop}%
\bibitem [{\citenamefont {Wilamowski}\ \emph {et~al.}(2008)\citenamefont
  {Wilamowski}, \citenamefont {Ungier},\ and\ \citenamefont
  {Jantsch}}]{wilamowski2008elec}%
  \BibitemOpen
  \bibfield  {author} {\bibinfo {author} {\bibfnamefont {Z.}~\bibnamefont
  {Wilamowski}}, \bibinfo {author} {\bibfnamefont {W.}~\bibnamefont {Ungier}},
  \ and\ \bibinfo {author} {\bibfnamefont {W.}~\bibnamefont {Jantsch}},\
  }\bibfield  {title} {\enquote {\bibinfo {title} {Electron {spin} {resonance}
  in a {two}-{dimensional} {electron} {gas} {induced} by {current} or by
  {electric} {field}},}\ }\href {\doibase 10.1103/PhysRevB.78.174423}
  {\bibfield  {journal} {\bibinfo  {journal} {Phys. Rev. B}\ }\textbf {\bibinfo
  {volume} {78}},\ \bibinfo {pages} {174423} (\bibinfo {year}
  {2008})}\BibitemShut {NoStop}%
\bibitem [{\citenamefont {Nadj-Perge}\ \emph {et~al.}(2010)\citenamefont
  {Nadj-Perge}, \citenamefont {Frolov}, \citenamefont {Bakkers},\ and\
  \citenamefont {Kouwenhoven}}]{nadj-perge2010qubit}%
  \BibitemOpen
  \bibfield  {author} {\bibinfo {author} {\bibfnamefont {S.}~\bibnamefont
  {Nadj-Perge}}, \bibinfo {author} {\bibfnamefont {S.~M.}\ \bibnamefont
  {Frolov}}, \bibinfo {author} {\bibfnamefont {P.~A.~M.}\ \bibnamefont
  {Bakkers}}, \ and\ \bibinfo {author} {\bibfnamefont {L.~P.}\ \bibnamefont
  {Kouwenhoven}},\ }\bibfield  {title} {\enquote {\bibinfo {title} {Spin-orbit
  qubit in a semiconductor nanowire},}\ }\href {\doibase 10.1038/nature09682}
  {\bibfield  {journal} {\bibinfo  {journal} {Nature}\ }\textbf {\bibinfo
  {volume} {468}},\ \bibinfo {pages} {1084} (\bibinfo {year}
  {2010})}\BibitemShut {NoStop}%
\bibitem [{\citenamefont {Stier}\ \emph {et~al.}(2023)\citenamefont {Stier},
  \citenamefont {Meining}, \citenamefont {Whiteside}, \citenamefont {McCombe},
  \citenamefont {Rashba}, \citenamefont {Grabs},\ and\ \citenamefont
  {Molenkamp}}]{Stier2023:PRB}%
  \BibitemOpen
  \bibfield  {author} {\bibinfo {author} {\bibfnamefont {A.~V.}\ \bibnamefont
  {Stier}}, \bibinfo {author} {\bibfnamefont {C.~J.}\ \bibnamefont {Meining}},
  \bibinfo {author} {\bibfnamefont {V.~R.}\ \bibnamefont {Whiteside}}, \bibinfo
  {author} {\bibfnamefont {B.~D.}\ \bibnamefont {McCombe}}, \bibinfo {author}
  {\bibfnamefont {E.~I.}\ \bibnamefont {Rashba}}, \bibinfo {author}
  {\bibfnamefont {P.}~\bibnamefont {Grabs}}, \ and\ \bibinfo {author}
  {\bibfnamefont {L.~W.}\ \bibnamefont {Molenkamp}},\ }\bibfield  {title}
  {\enquote {\bibinfo {title} {Electric-{dipole} {spin} {resonance} and
  {spin}-{orbit} {coupling} {effects} in {odd}-{integer} {quantum} {Hall}
  {edge} {channels}},}\ }\href {\doibase 10.1103/PhysRevB.107.045301}
  {\bibfield  {journal} {\bibinfo  {journal} {Phys. Rev. B}\ }\textbf {\bibinfo
  {volume} {107}},\ \bibinfo {pages} {045301} (\bibinfo {year}
  {2023})}\BibitemShut {NoStop}%
\bibitem [{\citenamefont {Brooks}\ and\ \citenamefont
  {Burkard}(2020)}]{Brooks2020:PRB}%
  \BibitemOpen
  \bibfield  {author} {\bibinfo {author} {\bibfnamefont {M.}~\bibnamefont
  {Brooks}}\ and\ \bibinfo {author} {\bibfnamefont {G.}~\bibnamefont
  {Burkard}},\ }\bibfield  {title} {\enquote {\bibinfo {title} {Electric
  {dipole} {spin} {resonance} of {two}-{dimensional} {semiconductor} {spin}
  {qubits}},}\ }\href {\doibase 10.1103/PhysRevB.101.035204} {\bibfield
  {journal} {\bibinfo  {journal} {Phys. Rev. B}\ }\textbf {\bibinfo {volume}
  {101}},\ \bibinfo {pages} {035204} (\bibinfo {year} {2020})}\BibitemShut
  {NoStop}%
\bibitem [{\citenamefont {Wang}\ \emph {et~al.}(2022)\citenamefont {Wang},
  \citenamefont {Xu}, \citenamefont {Gao}, \citenamefont {Liu}, \citenamefont
  {Ma}, \citenamefont {Zhang}, \citenamefont {Wang}, \citenamefont {Cao},
  \citenamefont {Wang}, \citenamefont {Zhang},\ and\ \citenamefont
  {et~al.}}]{Wang2022:NC}%
  \BibitemOpen
  \bibfield  {author} {\bibinfo {author} {\bibfnamefont {K.}~\bibnamefont
  {Wang}}, \bibinfo {author} {\bibfnamefont {G.}~\bibnamefont {Xu}}, \bibinfo
  {author} {\bibfnamefont {F.}~\bibnamefont {Gao}}, \bibinfo {author}
  {\bibfnamefont {He}~\bibnamefont {Liu}}, \bibinfo {author} {\bibfnamefont
  {R.-L.}\ \bibnamefont {Ma}}, \bibinfo {author} {\bibfnamefont
  {X.}~\bibnamefont {Zhang}}, \bibinfo {author} {\bibfnamefont
  {Z.}~\bibnamefont {Wang}}, \bibinfo {author} {\bibfnamefont {G.}~\bibnamefont
  {Cao}}, \bibinfo {author} {\bibfnamefont {T.}~\bibnamefont {Wang}}, \bibinfo
  {author} {\bibfnamefont {J.-J.}\ \bibnamefont {Zhang}}, \ and\ \bibinfo
  {author} {\bibnamefont {et~al.}},\ }\bibfield  {title} {\enquote {\bibinfo
  {title} {Ultrafast coherent control of a hole spin qubit in a germanium
  quantum dot},}\ }\href {\doibase 10.1038/s41467-021-27880-7} {\bibfield
  {journal} {\bibinfo  {journal} {Nat. Commun.}\ }\textbf {\bibinfo {volume}
  {13}},\ \bibinfo {pages} {206} (\bibinfo {year} {2022})}\BibitemShut
  {NoStop}%
\bibitem [{\citenamefont {Stano}\ and\ \citenamefont
  {Loss}(2022)}]{Stano2022:NRM}%
  \BibitemOpen
  \bibfield  {author} {\bibinfo {author} {\bibfnamefont {P.}~\bibnamefont
  {Stano}}\ and\ \bibinfo {author} {\bibfnamefont {D.}~\bibnamefont {Loss}},\
  }\bibfield  {title} {\enquote {\bibinfo {title} {Review of performance
  metrics of spin qubits in gated semiconducting nanostructures},}\ }\href
  {\doibase 10.1038/s42254-022-00484-w} {\bibfield  {journal} {\bibinfo
  {journal} {Nat. Rev. Mater.}\ }\textbf {\bibinfo {volume} {4}},\ \bibinfo
  {pages} {672} (\bibinfo {year} {2022})}\BibitemShut {NoStop}%
\bibitem [{\citenamefont {Tuan}\ \emph {et~al.}(2014)\citenamefont {Tuan},
  \citenamefont {Ortmann}, \citenamefont {Soriano}, \citenamefont
  {Valenzuela},\ and\ \citenamefont {Roche}}]{tuan2014pseudospin}%
  \BibitemOpen
  \bibfield  {author} {\bibinfo {author} {\bibfnamefont {D.~van}\ \bibnamefont
  {Tuan}}, \bibinfo {author} {\bibfnamefont {F.}~\bibnamefont {Ortmann}},
  \bibinfo {author} {\bibfnamefont {D.}~\bibnamefont {Soriano}}, \bibinfo
  {author} {\bibfnamefont {S.~O.}\ \bibnamefont {Valenzuela}}, \ and\ \bibinfo
  {author} {\bibfnamefont {S.}~\bibnamefont {Roche}},\ }\bibfield  {title}
  {\enquote {\bibinfo {title} {Pseudospin-driven spin relaxation mechanism in
  graphene},}\ }\href {\doibase 10.1038/NPHYS3083} {\bibfield  {journal}
  {\bibinfo  {journal} {Nat. Phys.}\ }\textbf {\bibinfo {volume} {10}},\
  \bibinfo {pages} {857} (\bibinfo {year} {2014})}\BibitemShut {NoStop}%
\bibitem [{\citenamefont {de~Moraes}\ \emph {et~al.}(2020)\citenamefont
  {de~Moraes}, \citenamefont {Cummings},\ and\ \citenamefont
  {Roche}}]{demoraes2020entanglement}%
  \BibitemOpen
  \bibfield  {author} {\bibinfo {author} {\bibfnamefont {B.~Gabrielly}\
  \bibnamefont {de~Moraes}}, \bibinfo {author} {\bibfnamefont {A.~W.}\
  \bibnamefont {Cummings}}, \ and\ \bibinfo {author} {\bibfnamefont
  {S.}~\bibnamefont {Roche}},\ }\bibfield  {title} {\enquote {\bibinfo {title}
  {Emergence of {intraparticle} {entanglement} and {time}-{varying} {violation}
  of {Bell}'s {inequality} in {Dirac} {matter}},}\ }\href {\doibase
  10.1103/PhysRevB.102.041403} {\bibfield  {journal} {\bibinfo  {journal}
  {Phys. Rev. B}\ }\textbf {\bibinfo {volume} {102}},\ \bibinfo {pages}
  {041403} (\bibinfo {year} {2020})}\BibitemShut {NoStop}%
\bibitem [{\citenamefont {Duckheim}\ and\ \citenamefont
  {Loss}(2007)}]{duckheim2007resonant}%
  \BibitemOpen
  \bibfield  {author} {\bibinfo {author} {\bibfnamefont {M.}~\bibnamefont
  {Duckheim}}\ and\ \bibinfo {author} {\bibfnamefont {D.}~\bibnamefont
  {Loss}},\ }\bibfield  {title} {\enquote {\bibinfo {title} {Resonant {spin}
  {polarization} and {spin} {current} in a {two}-{dimensional} {electron}
  {gas}},}\ }\href {\doibase 10.1103/PhysRevB.75.201305} {\bibfield  {journal}
  {\bibinfo  {journal} {Phys. Rev. B}\ }\textbf {\bibinfo {volume} {75}},\
  \bibinfo {pages} {201305} (\bibinfo {year} {2007})}\BibitemShut {NoStop}%
\bibitem [{\citenamefont {Wilamowski}\ \emph {et~al.}(2007)\citenamefont
  {Wilamowski}, \citenamefont {Malissa}, \citenamefont {Sch\"affler},\ and\
  \citenamefont {Jantsch}}]{wilamowski2007g}%
  \BibitemOpen
  \bibfield  {author} {\bibinfo {author} {\bibfnamefont {Z.}~\bibnamefont
  {Wilamowski}}, \bibinfo {author} {\bibfnamefont {H.}~\bibnamefont {Malissa}},
  \bibinfo {author} {\bibfnamefont {F.}~\bibnamefont {Sch\"affler}}, \ and\
  \bibinfo {author} {\bibfnamefont {W.}~\bibnamefont {Jantsch}},\ }\bibfield
  {title} {\enquote {\bibinfo {title} {$g$-{Factor} {Tuning} and {Manipulation}
  of {Spins} by an {Electric} {Current}},}\ }\href {\doibase
  10.1103/PhysRevLett.98.187203} {\bibfield  {journal} {\bibinfo  {journal}
  {Phys. Rev. Lett.}\ }\textbf {\bibinfo {volume} {98}},\ \bibinfo {pages}
  {187203} (\bibinfo {year} {2007})}\BibitemShut {NoStop}%
\bibitem [{\citenamefont {Weil}\ and\ \citenamefont
  {Bolton}(2007)}]{weil2007electron}%
  \BibitemOpen
  \bibfield  {author} {\bibinfo {author} {\bibfnamefont {J.~A.}\ \bibnamefont
  {Weil}}\ and\ \bibinfo {author} {\bibfnamefont {J.~R.}\ \bibnamefont
  {Bolton}},\ }\href@noop {} {\emph {\bibinfo {title} {Electron Paramagnetic
  Resonance: Elementary Theory and Practical Applications}}}\ (\bibinfo
  {publisher} {John Wiley \& Sons},\ \bibinfo {year} {2007})\BibitemShut
  {NoStop}%
\bibitem [{\citenamefont {Rashba}\ and\ \citenamefont
  {Efros}(2003)}]{rashba2003efficient}%
  \BibitemOpen
  \bibfield  {author} {\bibinfo {author} {\bibfnamefont {E.~I.}\ \bibnamefont
  {Rashba}}\ and\ \bibinfo {author} {\bibfnamefont {Al.~L.}\ \bibnamefont
  {Efros}},\ }\bibfield  {title} {\enquote {\bibinfo {title} {Efficient
  electron spin manipulation in a quantum well by an in-plane electric
  field},}\ }\href {\doibase 10.1063/1.1635987} {\bibfield  {journal} {\bibinfo
   {journal} {Appl. Phys. Lett.}\ }\textbf {\bibinfo {volume} {83}},\ \bibinfo
  {pages} {5295} (\bibinfo {year} {2003})}\BibitemShut {NoStop}%
\bibitem [{\citenamefont {Kim}(2017)}]{Kim2017}%
  \BibitemOpen
  \bibfield  {author} {\bibinfo {author} {\bibfnamefont {P.}~\bibnamefont
  {Kim}},\ }\enquote {\bibinfo {title} {{Graphene and Relativistic Quantum
  Physics}},}\ in\ \href@noop {} {\emph {\bibinfo {booktitle} {{Dirac
  Matter}}}},\ \bibinfo {editor} {edited by\ \bibinfo {editor} {\bibfnamefont
  {B.}~\bibnamefont {Duplantier}}, \bibinfo {editor} {\bibfnamefont
  {V.}~\bibnamefont {Rivasseau}}, \ and\ \bibinfo {editor} {\bibfnamefont
  {J.-N.}\ \bibnamefont {Fuchs}}}\ (\bibinfo  {publisher} {Springer
  International Publishing},\ \bibinfo {address} {Cham},\ \bibinfo {year}
  {2017})\ pp.\ \bibinfo {pages} {1--23}\BibitemShut {NoStop}%
\bibitem [{SM()}]{SM}%
  \BibitemOpen
  \href@noop {} {\enquote {\bibinfo {title} {{See Supplemental Material for
  expanded discussion of theoretical and computational methods. The
  Supplemental Material also contains Refs.[52,53]}},}\ }\BibitemShut {NoStop}%
\bibitem [{\citenamefont {Nomura}\ and\ \citenamefont
  {MacDonald}(2007)}]{nomura2007quantum}%
  \BibitemOpen
  \bibfield  {author} {\bibinfo {author} {\bibfnamefont {K.}~\bibnamefont
  {Nomura}}\ and\ \bibinfo {author} {\bibfnamefont {A.~H.}\ \bibnamefont
  {MacDonald}},\ }\bibfield  {title} {\enquote {\bibinfo {title} {Quantum
  transport of massless dirac fermions},}\ }\href {\doibase
  10.1103/PhysRevLett.98.076602} {\bibfield  {journal} {\bibinfo  {journal}
  {Phys. Rev. Lett.}\ }\textbf {\bibinfo {volume} {98}},\ \bibinfo {pages}
  {076602} (\bibinfo {year} {2007})}\BibitemShut {NoStop}%
\bibitem [{\citenamefont {Barati}\ and\ \citenamefont
  {Abedinpour}(2017)}]{barati2017optical}%
  \BibitemOpen
  \bibfield  {author} {\bibinfo {author} {\bibfnamefont {S.}~\bibnamefont
  {Barati}}\ and\ \bibinfo {author} {\bibfnamefont {S.~H.}\ \bibnamefont
  {Abedinpour}},\ }\bibfield  {title} {\enquote {\bibinfo {title} {Optical
  conductivity of three and two dimensional topological nodal-line
  semimetals},}\ }\href {\doibase 10.1103/PhysRevB.96.155150} {\bibfield
  {journal} {\bibinfo  {journal} {Phys. Rev. B}\ }\textbf {\bibinfo {volume}
  {96}},\ \bibinfo {pages} {155150} (\bibinfo {year} {2017})}\BibitemShut
  {NoStop}%
\bibitem [{\citenamefont {Sakurai}\ and\ \citenamefont
  {Napolitano}(2011)}]{sakurai1995modern}%
  \BibitemOpen
  \bibfield  {author} {\bibinfo {author} {\bibfnamefont {J.~J.}\ \bibnamefont
  {Sakurai}}\ and\ \bibinfo {author} {\bibfnamefont {J.}~\bibnamefont
  {Napolitano}},\ }\href@noop {} {\emph {\bibinfo {title} {Modern Quantum
  Mechanics, 2nd Edition}}}\ (\bibinfo  {publisher} {Addison-Wesley, San
  Francisco},\ \bibinfo {year} {2011})\ p.\ \bibinfo {pages} {342}\BibitemShut
  {NoStop}%
\bibitem [{\citenamefont {Griffiths}(2018)}]{griffiths2018introduction}%
  \BibitemOpen
  \bibfield  {author} {\bibinfo {author} {\bibfnamefont {D.~J.}\ \bibnamefont
  {Griffiths}},\ }\href@noop {} {\emph {\bibinfo {title} {Introduction to
  Quantum Mechanics,2nd Edition}}}\ (\bibinfo  {publisher} {Prentice Hall,
  Upper Saddle River, NJ},\ \bibinfo {year} {2018})\ p.\ \bibinfo {pages}
  {178}\BibitemShut {NoStop}%
\bibitem [{\citenamefont {Gmitra}\ \emph {et~al.}(2009)\citenamefont {Gmitra},
  \citenamefont {Konschuh}, \citenamefont {Ertler}, \citenamefont
  {Ambrosch-Draxl},\ and\ \citenamefont {Fabian}}]{gmitra2009band}%
  \BibitemOpen
  \bibfield  {author} {\bibinfo {author} {\bibfnamefont {M.}~\bibnamefont
  {Gmitra}}, \bibinfo {author} {\bibfnamefont {S.}~\bibnamefont {Konschuh}},
  \bibinfo {author} {\bibfnamefont {C.}~\bibnamefont {Ertler}}, \bibinfo
  {author} {\bibfnamefont {C.}~\bibnamefont {Ambrosch-Draxl}}, \ and\ \bibinfo
  {author} {\bibfnamefont {J.}~\bibnamefont {Fabian}},\ }\bibfield  {title}
  {\enquote {\bibinfo {title} {Band-{structure} {topologies} of {graphene}:
  {Spin}-{orbit} {coupling} {effects} from {first} {principles}},}\ }\href
  {\doibase 10.1103/PhysRevB.80.235431} {\bibfield  {journal} {\bibinfo
  {journal} {Phys. Rev. B}\ }\textbf {\bibinfo {volume} {80}},\ \bibinfo
  {pages} {235431} (\bibinfo {year} {2009})}\BibitemShut {NoStop}%
\bibitem [{\citenamefont {Min}\ \emph {et~al.}(2006)\citenamefont {Min},
  \citenamefont {Hill}, \citenamefont {Sinitsyn}, \citenamefont {Sahu},
  \citenamefont {Kleinman},\ and\ \citenamefont
  {MacDonald}}]{min2006intrinsic}%
  \BibitemOpen
  \bibfield  {author} {\bibinfo {author} {\bibfnamefont {H.}~\bibnamefont
  {Min}}, \bibinfo {author} {\bibfnamefont {J.~E.}\ \bibnamefont {Hill}},
  \bibinfo {author} {\bibfnamefont {N.~A.}\ \bibnamefont {Sinitsyn}}, \bibinfo
  {author} {\bibfnamefont {B.~R.}\ \bibnamefont {Sahu}}, \bibinfo {author}
  {\bibfnamefont {L.}~\bibnamefont {Kleinman}}, \ and\ \bibinfo {author}
  {\bibfnamefont {A.~H.}\ \bibnamefont {MacDonald}},\ }\bibfield  {title}
  {\enquote {\bibinfo {title} {Intrinsic and {Rashba} {spin}-{orbit}
  {interactions} in {graphene} {sheets}},}\ }\href {\doibase
  10.1103/PhysRevB.74.165310} {\bibfield  {journal} {\bibinfo  {journal} {Phys.
  Rev. B}\ }\textbf {\bibinfo {volume} {74}},\ \bibinfo {pages} {165310}
  (\bibinfo {year} {2006})}\BibitemShut {NoStop}%
\bibitem [{\citenamefont {Cao}\ \emph {et~al.}(2012)\citenamefont {Cao},
  \citenamefont {Wang}, \citenamefont {Han}, \citenamefont {Ye}, \citenamefont
  {Zhu}, \citenamefont {Shi}, \citenamefont {Niu}, \citenamefont {Tan},
  \citenamefont {Wang},\ and\ \citenamefont {Liu}}]{cao2012valley}%
  \BibitemOpen
  \bibfield  {author} {\bibinfo {author} {\bibfnamefont {T.}~\bibnamefont
  {Cao}}, \bibinfo {author} {\bibfnamefont {G.}~\bibnamefont {Wang}}, \bibinfo
  {author} {\bibfnamefont {W.}~\bibnamefont {Han}}, \bibinfo {author}
  {\bibfnamefont {H.}~\bibnamefont {Ye}}, \bibinfo {author} {\bibfnamefont
  {C.}~\bibnamefont {Zhu}}, \bibinfo {author} {\bibfnamefont {J.}~\bibnamefont
  {Shi}}, \bibinfo {author} {\bibfnamefont {Q.}~\bibnamefont {Niu}}, \bibinfo
  {author} {\bibfnamefont {P.}~\bibnamefont {Tan}}, \bibinfo {author}
  {\bibfnamefont {E.}~\bibnamefont {Wang}}, \ and\ \bibinfo {author}
  {\bibfnamefont {B.}~\bibnamefont {Liu}},\ }\bibfield  {title} {\enquote
  {\bibinfo {title} {Valley-selective circular dichroism of monolayer
  molybdenum disulphide},}\ }\href {\doibase DOI: 10.1038/ncomms1882}
  {\bibfield  {journal} {\bibinfo  {journal} {Nat. Comm.}\ }\textbf {\bibinfo
  {volume} {3}},\ \bibinfo {pages} {887} (\bibinfo {year} {2012})}\BibitemShut
  {NoStop}%
\bibitem [{\citenamefont {Mak}\ \emph {et~al.}(2012)\citenamefont {Mak},
  \citenamefont {He}, \citenamefont {Shan},\ and\ \citenamefont
  {Heinz}}]{mak2012control}%
  \BibitemOpen
  \bibfield  {author} {\bibinfo {author} {\bibfnamefont {K.~F.}\ \bibnamefont
  {Mak}}, \bibinfo {author} {\bibfnamefont {K.}~\bibnamefont {He}}, \bibinfo
  {author} {\bibfnamefont {J.}~\bibnamefont {Shan}}, \ and\ \bibinfo {author}
  {\bibfnamefont {T.~F.}\ \bibnamefont {Heinz}},\ }\bibfield  {title} {\enquote
  {\bibinfo {title} {Control of valley polarization in monolayer {MoS$_2$} by
  optical helicity},}\ }\href {\doibase 10.1038/NNANO.2012.96} {\bibfield
  {journal} {\bibinfo  {journal} {Nat. Nanotechnol.}\ }\textbf {\bibinfo
  {volume} {7}},\ \bibinfo {pages} {494} (\bibinfo {year} {2012})}\BibitemShut
  {NoStop}%
\bibitem [{\citenamefont {Xiao}\ \emph {et~al.}(2012)\citenamefont {Xiao},
  \citenamefont {Liu}, \citenamefont {Feng}, \citenamefont {Xu},\ and\
  \citenamefont {Yao}}]{xiao2012coupled}%
  \BibitemOpen
  \bibfield  {author} {\bibinfo {author} {\bibfnamefont {D.}~\bibnamefont
  {Xiao}}, \bibinfo {author} {\bibfnamefont {G.}~\bibnamefont {Liu}}, \bibinfo
  {author} {\bibfnamefont {W.}~\bibnamefont {Feng}}, \bibinfo {author}
  {\bibfnamefont {X.}~\bibnamefont {Xu}}, \ and\ \bibinfo {author}
  {\bibfnamefont {W.}~\bibnamefont {Yao}},\ }\bibfield  {title} {\enquote
  {\bibinfo {title} {Coupled {Spin} and {Valley} {Physics} in {Monolayers} of
  {MoS$_2$} and {Other Group}-{VI} {Dichalcogenides}},}\ }\href {\doibase
  10.1103/PhysRevLett.108.196802} {\bibfield  {journal} {\bibinfo  {journal}
  {Phys. Rev. Lett.}\ }\textbf {\bibinfo {volume} {108}},\ \bibinfo {pages}
  {196802} (\bibinfo {year} {2012})}\BibitemShut {NoStop}%
\bibitem [{\citenamefont {Inglot}\ \emph {et~al.}(2014)\citenamefont {Inglot},
  \citenamefont {Dugaev}, \citenamefont {Sherman},\ and\ \citenamefont
  {Barna\ifmmode~\acute{s}\else \'{s}\fi{}}}]{inglot2014optical}%
  \BibitemOpen
  \bibfield  {author} {\bibinfo {author} {\bibfnamefont {M.}~\bibnamefont
  {Inglot}}, \bibinfo {author} {\bibfnamefont {V.~K.}\ \bibnamefont {Dugaev}},
  \bibinfo {author} {\bibfnamefont {E.~Ya.}\ \bibnamefont {Sherman}}, \ and\
  \bibinfo {author} {\bibfnamefont {J.}~\bibnamefont
  {Barna\ifmmode~\acute{s}\else \'{s}\fi{}}},\ }\bibfield  {title} {\enquote
  {\bibinfo {title} {Optical {spin} {injection} in {graphene} with {Rashba}
  {spin}-{orbit} {interaction}},}\ }\href {\doibase 10.1103/PhysRevB.89.155411}
  {\bibfield  {journal} {\bibinfo  {journal} {Phys. Rev. B}\ }\textbf {\bibinfo
  {volume} {89}},\ \bibinfo {pages} {155411} (\bibinfo {year}
  {2014})}\BibitemShut {NoStop}%
\bibitem [{\citenamefont {Kumar}\ \emph {et~al.}(2021)\citenamefont {Kumar},
  \citenamefont {Maiti},\ and\ \citenamefont {Maslov}}]{kumar2021zero}%
  \BibitemOpen
  \bibfield  {author} {\bibinfo {author} {\bibfnamefont {A.}~\bibnamefont
  {Kumar}}, \bibinfo {author} {\bibfnamefont {S.}~\bibnamefont {Maiti}}, \ and\
  \bibinfo {author} {\bibfnamefont {D.~L.}\ \bibnamefont {Maslov}},\ }\bibfield
   {title} {\enquote {\bibinfo {title} {Zero-{field} {spin} {resonance} in
  {graphene} with {proximity}-{induced} {spin}-{orbit} {coupling}},}\ }\href
  {\doibase 10.1103/PhysRevB.104.155138} {\bibfield  {journal} {\bibinfo
  {journal} {Phys. Rev. B}\ }\textbf {\bibinfo {volume} {104}},\ \bibinfo
  {pages} {155138} (\bibinfo {year} {2021})}\BibitemShut {NoStop}%
\bibitem [{\citenamefont {Offidani}\ \emph {et~al.}(2018)\citenamefont
  {Offidani}, \citenamefont {Raimondi},\ and\ \citenamefont
  {Ferreira}}]{offidani2018microscopic}%
  \BibitemOpen
  \bibfield  {author} {\bibinfo {author} {\bibfnamefont {M.}~\bibnamefont
  {Offidani}}, \bibinfo {author} {\bibfnamefont {R.}~\bibnamefont {Raimondi}},
  \ and\ \bibinfo {author} {\bibfnamefont {A.}~\bibnamefont {Ferreira}},\
  }\bibfield  {title} {\enquote {\bibinfo {title} {Microscopic {Linear}
  {Response} {Theory} of {Spin} {Relaxation} and {Relativistic} {Transport}
  {Phenomena} in {Graphene}},}\ }\href {\doibase 10.3390/condmat3020018}
  {\bibfield  {journal} {\bibinfo  {journal} {Condens. Matter}\ }\textbf
  {\bibinfo {volume} {3}},\ \bibinfo {pages} {18} (\bibinfo {year}
  {2018})}\BibitemShut {NoStop}%
\bibitem [{\citenamefont {Wu}\ \emph {et~al.}(2010)\citenamefont {Wu},
  \citenamefont {Jiang},\ and\ \citenamefont {Weng}}]{wu2010spin}%
  \BibitemOpen
  \bibfield  {author} {\bibinfo {author} {\bibfnamefont {M.~W.}\ \bibnamefont
  {Wu}}, \bibinfo {author} {\bibfnamefont {J.~H.}\ \bibnamefont {Jiang}}, \
  and\ \bibinfo {author} {\bibfnamefont {M.~Q.}\ \bibnamefont {Weng}},\
  }\bibfield  {title} {\enquote {\bibinfo {title} {Spin dynamics in
  semiconductors},}\ }\href {\doibase 10.1016/j.physrep.2010.04.002} {\bibfield
   {journal} {\bibinfo  {journal} {Phys. Rep.}\ }\textbf {\bibinfo {volume}
  {493}},\ \bibinfo {pages} {61} (\bibinfo {year} {2010})}\BibitemShut
  {NoStop}%
\bibitem [{\citenamefont {Raes}\ \emph {et~al.}(2016)\citenamefont {Raes},
  \citenamefont {Scheerder}, \citenamefont {Costache}, \citenamefont {Bonell},
  \citenamefont {Sierra}, \citenamefont {Cuppens}, \citenamefont {Van~de
  Vondel},\ and\ \citenamefont {Valenzuela}}]{raes2016determination}%
  \BibitemOpen
  \bibfield  {author} {\bibinfo {author} {\bibfnamefont {B.}~\bibnamefont
  {Raes}}, \bibinfo {author} {\bibfnamefont {J.~E.}\ \bibnamefont {Scheerder}},
  \bibinfo {author} {\bibfnamefont {M.~V.}\ \bibnamefont {Costache}}, \bibinfo
  {author} {\bibfnamefont {F.}~\bibnamefont {Bonell}}, \bibinfo {author}
  {\bibfnamefont {J.~F.}\ \bibnamefont {Sierra}}, \bibinfo {author}
  {\bibfnamefont {J.}~\bibnamefont {Cuppens}}, \bibinfo {author} {\bibfnamefont
  {J.}~\bibnamefont {Van~de Vondel}}, \ and\ \bibinfo {author} {\bibfnamefont
  {S.~O.}\ \bibnamefont {Valenzuela}},\ }\bibfield  {title} {\enquote {\bibinfo
  {title} {Determination of the spin-lifetime anisotropy in graphene using
  oblique spin precession},}\ }\href {\doibase 10.1038/ncomms11444} {\bibfield
  {journal} {\bibinfo  {journal} {Nat. Commun.}\ }\textbf {\bibinfo {volume}
  {7}},\ \bibinfo {pages} {11444} (\bibinfo {year} {2016})}\BibitemShut
  {NoStop}%
\bibitem [{\citenamefont {Raes}\ \emph {et~al.}(2017)\citenamefont {Raes},
  \citenamefont {Cummings}, \citenamefont {Bonell}, \citenamefont {Costache},
  \citenamefont {Sierra}, \citenamefont {Roche},\ and\ \citenamefont
  {Valenzuela}}]{raes2017anisotropy}%
  \BibitemOpen
  \bibfield  {author} {\bibinfo {author} {\bibfnamefont {B.}~\bibnamefont
  {Raes}}, \bibinfo {author} {\bibfnamefont {A.~W.}\ \bibnamefont {Cummings}},
  \bibinfo {author} {\bibfnamefont {F.}~\bibnamefont {Bonell}}, \bibinfo
  {author} {\bibfnamefont {M.~V.}\ \bibnamefont {Costache}}, \bibinfo {author}
  {\bibfnamefont {J.~F.}\ \bibnamefont {Sierra}}, \bibinfo {author}
  {\bibfnamefont {S.}~\bibnamefont {Roche}}, \ and\ \bibinfo {author}
  {\bibfnamefont {S.~O.}\ \bibnamefont {Valenzuela}},\ }\bibfield  {title}
  {\enquote {\bibinfo {title} {Spin {precession} in {anisotropic} {media}},}\
  }\href {\doibase 10.1103/PhysRevB.95.085403} {\bibfield  {journal} {\bibinfo
  {journal} {Phys. Rev B}\ }\textbf {\bibinfo {volume} {95}},\ \bibinfo {pages}
  {085403} (\bibinfo {year} {2017})}\BibitemShut {NoStop}%
\bibitem [{\citenamefont {Silsbee}(1980)}]{silsbee1980}%
  \BibitemOpen
  \bibfield  {author} {\bibinfo {author} {\bibfnamefont {R.~H.}\ \bibnamefont
  {Silsbee}},\ }\bibfield  {title} {\enquote {\bibinfo {title} {Novel method
  for the study of spin transport in conductors},}\ }\href@noop {} {\bibfield
  {journal} {\bibinfo  {journal} {Bull. Magn. Reson.}\ }\textbf {\bibinfo
  {volume} {2}},\ \bibinfo {pages} {284} (\bibinfo {year} {1980})}\BibitemShut
  {NoStop}%
\bibitem [{\citenamefont {Johnson}\ and\ \citenamefont
  {Silsbee}(1985)}]{johnson1985Interfacial}%
  \BibitemOpen
  \bibfield  {author} {\bibinfo {author} {\bibfnamefont {M.}~\bibnamefont
  {Johnson}}\ and\ \bibinfo {author} {\bibfnamefont {R.~H.}\ \bibnamefont
  {Silsbee}},\ }\bibfield  {title} {\enquote {\bibinfo {title} {Interfacial
  {Charge}-{Spin} {Coupling}: {Injection} and {Detection} of {Spin}
  {Magnetization} in {Metals}},}\ }\href {\doibase 10.1103/PhysRevLett.55.1790}
  {\bibfield  {journal} {\bibinfo  {journal} {Phys. Rev. Lett.}\ }\textbf
  {\bibinfo {volume} {55}},\ \bibinfo {pages} {1790} (\bibinfo {year}
  {1985})}\BibitemShut {NoStop}%
\bibitem [{\citenamefont {{\v{Z}uti\'c}}\ \emph {et~al.}(2002)\citenamefont
  {{\v{Z}uti\'c}}, \citenamefont {Fabian},\ and\ \citenamefont {{Das
  Sarma}}}]{Zutic2002:PRL}%
  \BibitemOpen
  \bibfield  {author} {\bibinfo {author} {\bibfnamefont {I.}~\bibnamefont
  {{\v{Z}uti\'c}}}, \bibinfo {author} {\bibfnamefont {J.}~\bibnamefont
  {Fabian}}, \ and\ \bibinfo {author} {\bibfnamefont {S.}~\bibnamefont {{Das
  Sarma}}},\ }\bibfield  {title} {\enquote {\bibinfo {title} {Spin-{Polarized}
  {Transport} in {Inhomogeneous} {Magnetic} {Semiconductors}: {Theory} of
  {Magnetic}/{Nonmagnetic} {\it p-n} {Junctions}},}\ }\href {\doibase
  10.1103/PhysRevLett.88.066603} {\bibfield  {journal} {\bibinfo  {journal}
  {Phys. Rev. Lett.}\ }\textbf {\bibinfo {volume} {88}},\ \bibinfo {pages}
  {066603} (\bibinfo {year} {2002})}\BibitemShut {NoStop}%
\bibitem [{\citenamefont {Fabian}\ \emph {et~al.}(2007)\citenamefont {Fabian},
  \citenamefont {Matos-Abiague}, \citenamefont {Ertler}, \citenamefont
  {Stano},\ and\ \citenamefont {{\v{Z}}uti{\'c}}}]{fabian2007semiconductor}%
  \BibitemOpen
  \bibfield  {author} {\bibinfo {author} {\bibfnamefont {J.}~\bibnamefont
  {Fabian}}, \bibinfo {author} {\bibfnamefont {A.}~\bibnamefont
  {Matos-Abiague}}, \bibinfo {author} {\bibfnamefont {C.}~\bibnamefont
  {Ertler}}, \bibinfo {author} {\bibfnamefont {P.}~\bibnamefont {Stano}}, \
  and\ \bibinfo {author} {\bibfnamefont {I.}~\bibnamefont {{\v{Z}}uti{\'c}}},\
  }\bibfield  {title} {\enquote {\bibinfo {title} {Semiconductor
  spintronics},}\ }\href@noop {} {\bibfield  {journal} {\bibinfo  {journal}
  {Acta Phys. Slovaca}\ }\textbf {\bibinfo {volume} {57}},\ \bibinfo {pages}
  {565} (\bibinfo {year} {2007})}\BibitemShut {NoStop}%
\bibitem [{\citenamefont {Xu}\ \emph {et~al.}(2018)\citenamefont {Xu},
  \citenamefont {Singh}, \citenamefont {Katoch}, \citenamefont {Wu},
  \citenamefont {Zhu}, \citenamefont {{\v{Z}}uti{\'c}},\ and\ \citenamefont
  {Kawakami}}]{xu2018spin}%
  \BibitemOpen
  \bibfield  {author} {\bibinfo {author} {\bibfnamefont {J.}~\bibnamefont
  {Xu}}, \bibinfo {author} {\bibfnamefont {S.}~\bibnamefont {Singh}}, \bibinfo
  {author} {\bibfnamefont {J.}~\bibnamefont {Katoch}}, \bibinfo {author}
  {\bibfnamefont {G.}~\bibnamefont {Wu}}, \bibinfo {author} {\bibfnamefont
  {T.}~\bibnamefont {Zhu}}, \bibinfo {author} {\bibfnamefont {I.}~\bibnamefont
  {{\v{Z}}uti{\'c}}}, \ and\ \bibinfo {author} {\bibfnamefont {R.~K.}\
  \bibnamefont {Kawakami}},\ }\bibfield  {title} {\enquote {\bibinfo {title}
  {Spin inversion in graphene spin valves by gate-tunable magnetic proximity
  effect at one-dimensional contacts},}\ }\href {\doibase
  10.1038/s41467-018-05358-3} {\bibfield  {journal} {\bibinfo  {journal} {Nat.
  Commun.}\ }\textbf {\bibinfo {volume} {9}},\ \bibinfo {pages} {2869}
  (\bibinfo {year} {2018})}\BibitemShut {NoStop}%
\bibitem [{\citenamefont {Dahal}\ and\ \citenamefont
  {Batzill}(2014)}]{dahal2014graphene}%
  \BibitemOpen
  \bibfield  {author} {\bibinfo {author} {\bibfnamefont {A.}~\bibnamefont
  {Dahal}}\ and\ \bibinfo {author} {\bibfnamefont {M.}~\bibnamefont
  {Batzill}},\ }\bibfield  {title} {\enquote {\bibinfo {title} {Graphene-nickel
  interfaces: a review},}\ }\href {\doibase 10.1039/C3NR05279F} {\bibfield
  {journal} {\bibinfo  {journal} {Nanoscale}\ }\textbf {\bibinfo {volume}
  {6}},\ \bibinfo {pages} {2548} (\bibinfo {year} {2014})}\BibitemShut
  {NoStop}%
\bibitem [{\citenamefont {Cattelan}\ \emph {et~al.}(2015)\citenamefont
  {Cattelan}, \citenamefont {Peng}, \citenamefont {Cavaliere}, \citenamefont
  {Artiglia}, \citenamefont {Barinov}, \citenamefont {Roling}, \citenamefont
  {Favaro}, \citenamefont {Pis}, \citenamefont {Nappini}, \citenamefont
  {Magnano},\ and\ \citenamefont {et~al.}}]{cattelan2015nature}%
  \BibitemOpen
  \bibfield  {author} {\bibinfo {author} {\bibfnamefont {M.}~\bibnamefont
  {Cattelan}}, \bibinfo {author} {\bibfnamefont {G.~W.}\ \bibnamefont {Peng}},
  \bibinfo {author} {\bibfnamefont {E.}~\bibnamefont {Cavaliere}}, \bibinfo
  {author} {\bibfnamefont {L.}~\bibnamefont {Artiglia}}, \bibinfo {author}
  {\bibfnamefont {A.}~\bibnamefont {Barinov}}, \bibinfo {author} {\bibfnamefont
  {L.~T.}\ \bibnamefont {Roling}}, \bibinfo {author} {\bibfnamefont
  {M.}~\bibnamefont {Favaro}}, \bibinfo {author} {\bibfnamefont
  {I.}~\bibnamefont {Pis}}, \bibinfo {author} {\bibfnamefont {S.}~\bibnamefont
  {Nappini}}, \bibinfo {author} {\bibfnamefont {E.}~\bibnamefont {Magnano}}, \
  and\ \bibinfo {author} {\bibnamefont {et~al.}},\ }\bibfield  {title}
  {\enquote {\bibinfo {title} {The nature of the {Fe}-graphene interface at the
  nanometer level},}\ }\href {\doibase 10.1039/C4NR04956J} {\bibfield
  {journal} {\bibinfo  {journal} {Nanoscale}\ }\textbf {\bibinfo {volume}
  {7}},\ \bibinfo {pages} {2450} (\bibinfo {year} {2015})}\BibitemShut
  {NoStop}%
\bibitem [{\citenamefont {Liu}\ \emph {et~al.}(2015)\citenamefont {Liu},
  \citenamefont {Wang}, \citenamefont {Wang}, \citenamefont {Wang},
  \citenamefont {Lu}, \citenamefont {Jin}, \citenamefont {Zhang}, \citenamefont
  {Zhang}, \citenamefont {Laan}, \citenamefont {Xu},\ and\ \citenamefont
  {et~al.}}]{liu2015atomic}%
  \BibitemOpen
  \bibfield  {author} {\bibinfo {author} {\bibfnamefont {W.~Q.}\ \bibnamefont
  {Liu}}, \bibinfo {author} {\bibfnamefont {W.~Y.}\ \bibnamefont {Wang}},
  \bibinfo {author} {\bibfnamefont {J.~J.}\ \bibnamefont {Wang}}, \bibinfo
  {author} {\bibfnamefont {F.~Q.}\ \bibnamefont {Wang}}, \bibinfo {author}
  {\bibfnamefont {C.}~\bibnamefont {Lu}}, \bibinfo {author} {\bibfnamefont
  {F.}~\bibnamefont {Jin}}, \bibinfo {author} {\bibfnamefont {A.}~\bibnamefont
  {Zhang}}, \bibinfo {author} {\bibfnamefont {Q.~M.}\ \bibnamefont {Zhang}},
  \bibinfo {author} {\bibfnamefont {G.}~\bibnamefont {Laan}}, \bibinfo {author}
  {\bibfnamefont {Y.~B.}\ \bibnamefont {Xu}}, \ and\ \bibinfo {author}
  {\bibnamefont {et~al.}},\ }\bibfield  {title} {\enquote {\bibinfo {title}
  {Atomic-{Scale} {Interfacial} {Magnetism} in {Fe}/{Graphene}
  {Heterojunction}},}\ }\href {\doibase 10.1038/srep11911 (2015)} {\bibfield
  {journal} {\bibinfo  {journal} {Sci. Rep.}\ }\textbf {\bibinfo {volume}
  {5}},\ \bibinfo {pages} {11911} (\bibinfo {year} {2015})}\BibitemShut
  {NoStop}%
\bibitem [{\citenamefont {Swartz}\ \emph {et~al.}(2012)\citenamefont {Swartz},
  \citenamefont {Odenthal}, \citenamefont {Hao}, \citenamefont {Ruoff},\ and\
  \citenamefont {Kawakami}}]{swartz2012integration}%
  \BibitemOpen
  \bibfield  {author} {\bibinfo {author} {\bibfnamefont {A.~G.}\ \bibnamefont
  {Swartz}}, \bibinfo {author} {\bibfnamefont {P.~M.}\ \bibnamefont
  {Odenthal}}, \bibinfo {author} {\bibfnamefont {Y.}~\bibnamefont {Hao}},
  \bibinfo {author} {\bibfnamefont {R.~S.}\ \bibnamefont {Ruoff}}, \ and\
  \bibinfo {author} {\bibfnamefont {R.~K.}\ \bibnamefont {Kawakami}},\
  }\bibfield  {title} {\enquote {\bibinfo {title} {Integration of the
  {Ferromagnetic} {Insulator} {EuO} onto {Graphene}},}\ }\href {\doibase
  10.1021/nn303771f} {\bibfield  {journal} {\bibinfo  {journal} {ACS Nano}\
  }\textbf {\bibinfo {volume} {6}},\ \bibinfo {pages} {10063} (\bibinfo {year}
  {2012})}\BibitemShut {NoStop}%
\bibitem [{\citenamefont {Wang}\ \emph {et~al.}(2015)\citenamefont {Wang},
  \citenamefont {Tang}, \citenamefont {Sachs}, \citenamefont {Barlas},\ and\
  \citenamefont {Shi}}]{wang2015proximity}%
  \BibitemOpen
  \bibfield  {author} {\bibinfo {author} {\bibfnamefont {Z.}~\bibnamefont
  {Wang}}, \bibinfo {author} {\bibfnamefont {C.}~\bibnamefont {Tang}}, \bibinfo
  {author} {\bibfnamefont {R.}~\bibnamefont {Sachs}}, \bibinfo {author}
  {\bibfnamefont {Y.}~\bibnamefont {Barlas}}, \ and\ \bibinfo {author}
  {\bibfnamefont {J.}~\bibnamefont {Shi}},\ }\bibfield  {title} {\enquote
  {\bibinfo {title} {Proximity-{Induced} {Ferromagnetism} in {Graphene}
  {Revealed} by the {Anomalous} {Hall} {Effect}},}\ }\href {\doibase
  10.1103/PhysRevLett.114.016603} {\bibfield  {journal} {\bibinfo  {journal}
  {Phys. Rev. Lett.}\ }\textbf {\bibinfo {volume} {114}},\ \bibinfo {pages}
  {016603} (\bibinfo {year} {2015})}\BibitemShut {NoStop}%
\bibitem [{\citenamefont {Zollner}\ \emph {et~al.}(2020)\citenamefont
  {Zollner}, \citenamefont {Petrovi{\'c}}, \citenamefont {Dolui}, \citenamefont
  {Plech{\'a}{\v{c}}}, \citenamefont {Nikoli{\'c}},\ and\ \citenamefont
  {Fabian}}]{zollner2020scattering}%
  \BibitemOpen
  \bibfield  {author} {\bibinfo {author} {\bibfnamefont {K.}~\bibnamefont
  {Zollner}}, \bibinfo {author} {\bibfnamefont {M.~D.}\ \bibnamefont
  {Petrovi{\'c}}}, \bibinfo {author} {\bibfnamefont {K.}~\bibnamefont {Dolui}},
  \bibinfo {author} {\bibfnamefont {P.}~\bibnamefont {Plech{\'a}{\v{c}}}},
  \bibinfo {author} {\bibfnamefont {B.~K.}\ \bibnamefont {Nikoli{\'c}}}, \ and\
  \bibinfo {author} {\bibfnamefont {J.}~\bibnamefont {Fabian}},\ }\bibfield
  {title} {\enquote {\bibinfo {title} {Scattering-{Induced} and {highly}
  {tunable} by {gate} {damping}-{like} {spin}-{orbit} {torque} in {graphene}
  {doubly} {proximitized} by {two}-{dimensional} {magnet} {Cr$_2$Ge$_2$Te$_6$}
  and {monolayer} {WS$_2$}},}\ }\href {\doibase
  10.1103/PhysRevResearch.2.043057} {\bibfield  {journal} {\bibinfo  {journal}
  {Phys. Rev. Res.}\ }\textbf {\bibinfo {volume} {2}},\ \bibinfo {pages}
  {043057} (\bibinfo {year} {2020})}\BibitemShut {NoStop}%
\bibitem [{\citenamefont {Tang}\ \emph {et~al.}(2020)\citenamefont {Tang},
  \citenamefont {Zhang}, \citenamefont {Lai}, \citenamefont {Tan},\ and\
  \citenamefont {Gao}}]{tang2020magnetic}%
  \BibitemOpen
  \bibfield  {author} {\bibinfo {author} {\bibfnamefont {C.}~\bibnamefont
  {Tang}}, \bibinfo {author} {\bibfnamefont {Z.}~\bibnamefont {Zhang}},
  \bibinfo {author} {\bibfnamefont {S.}~\bibnamefont {Lai}}, \bibinfo {author}
  {\bibfnamefont {Q.}~\bibnamefont {Tan}}, \ and\ \bibinfo {author}
  {\bibfnamefont {W.}~\bibnamefont {Gao}},\ }\bibfield  {title} {\enquote
  {\bibinfo {title} {Magnetic {Proximity} {Effect} in {Graphene}/{CrBr$_3$} van
  der {Waals} {Heterostructures}},}\ }\href {\doibase 10.1002/adma.201908498}
  {\bibfield  {journal} {\bibinfo  {journal} {Adv. Mater.}\ }\textbf {\bibinfo
  {volume} {32}},\ \bibinfo {pages} {1908498} (\bibinfo {year}
  {2020})}\BibitemShut {NoStop}%
\bibitem [{\citenamefont {Zhou}\ \emph {et~al.}(2018)\citenamefont {Zhou},
  \citenamefont {Ji}, \citenamefont {Tian}, \citenamefont {Cheng},
  \citenamefont {Wang},\ and\ \citenamefont {Mi}}]{zhou2018proximity}%
  \BibitemOpen
  \bibfield  {author} {\bibinfo {author} {\bibfnamefont {B.}~\bibnamefont
  {Zhou}}, \bibinfo {author} {\bibfnamefont {S.}~\bibnamefont {Ji}}, \bibinfo
  {author} {\bibfnamefont {Z.}~\bibnamefont {Tian}}, \bibinfo {author}
  {\bibfnamefont {W.}~\bibnamefont {Cheng}}, \bibinfo {author} {\bibfnamefont
  {X.}~\bibnamefont {Wang}}, \ and\ \bibinfo {author} {\bibfnamefont
  {W.}~\bibnamefont {Mi}},\ }\bibfield  {title} {\enquote {\bibinfo {title}
  {Proximity effect induced spin filtering and gap opening in graphene by
  half-metallic monolayer {Cr$_2$C} ferromagnet},}\ }\href {\doibase
  10.1016/j.carbon.2018.02.044} {\bibfield  {journal} {\bibinfo  {journal}
  {Carbon}\ }\textbf {\bibinfo {volume} {132}},\ \bibinfo {pages} {25}
  (\bibinfo {year} {2018})}\BibitemShut {NoStop}%
\bibitem [{\citenamefont {Farooq}\ and\ \citenamefont
  {Hong}(2019)}]{farooq2019switchable}%
  \BibitemOpen
  \bibfield  {author} {\bibinfo {author} {\bibfnamefont {M.~U.}\ \bibnamefont
  {Farooq}}\ and\ \bibinfo {author} {\bibfnamefont {J.}~\bibnamefont {Hong}},\
  }\bibfield  {title} {\enquote {\bibinfo {title} {Switchable valley splitting
  by external electric field effect in graphene/{CrI$_3$} heterostructures},}\
  }\href {\doibase 10.1038/s41699-019-0086-6} {\bibfield  {journal} {\bibinfo
  {journal} {npj 2D Mater. Appl.}\ }\textbf {\bibinfo {volume} {3}},\ \bibinfo
  {pages} {3} (\bibinfo {year} {2019})}\BibitemShut {NoStop}%
\bibitem [{\citenamefont {Kara}\ \emph {et~al.}(2012)\citenamefont {Kara},
  \citenamefont {Enriquez}, \citenamefont {Seitsonen}, \citenamefont {Voon},
  \citenamefont {Vizzini}, \citenamefont {Aufray},\ and\ \citenamefont
  {Oughaddou}}]{kara2012review}%
  \BibitemOpen
  \bibfield  {author} {\bibinfo {author} {\bibfnamefont {A.}~\bibnamefont
  {Kara}}, \bibinfo {author} {\bibfnamefont {H.}~\bibnamefont {Enriquez}},
  \bibinfo {author} {\bibfnamefont {A.~P.}\ \bibnamefont {Seitsonen}}, \bibinfo
  {author} {\bibfnamefont {L.~C. Lew~Yan}\ \bibnamefont {Voon}}, \bibinfo
  {author} {\bibfnamefont {S.}~\bibnamefont {Vizzini}}, \bibinfo {author}
  {\bibfnamefont {B.}~\bibnamefont {Aufray}}, \ and\ \bibinfo {author}
  {\bibfnamefont {H.}~\bibnamefont {Oughaddou}},\ }\bibfield  {title} {\enquote
  {\bibinfo {title} {A review on silicene $-${New} candidate for
  electronics},}\ }\href {\doibase 10.1016/j.surfrep.2011.10.001} {\bibfield
  {journal} {\bibinfo  {journal} {Surf. Sci. Rep.}\ }\textbf {\bibinfo {volume}
  {67}},\ \bibinfo {pages} {1} (\bibinfo {year} {2012})}\BibitemShut {NoStop}%
\bibitem [{\citenamefont {Reis}\ \emph {et~al.}(2017)\citenamefont {Reis},
  \citenamefont {Li}, \citenamefont {Dudy}, \citenamefont {Bauernfeind},
  \citenamefont {Glass}, \citenamefont {Hanke}, \citenamefont {Thomale},
  \citenamefont {Sch{\"a}fer},\ and\ \citenamefont
  {Claessen}}]{reis2017bismuthene}%
  \BibitemOpen
  \bibfield  {author} {\bibinfo {author} {\bibfnamefont {F.}~\bibnamefont
  {Reis}}, \bibinfo {author} {\bibfnamefont {G.}~\bibnamefont {Li}}, \bibinfo
  {author} {\bibfnamefont {L.}~\bibnamefont {Dudy}}, \bibinfo {author}
  {\bibfnamefont {M.}~\bibnamefont {Bauernfeind}}, \bibinfo {author}
  {\bibfnamefont {S.}~\bibnamefont {Glass}}, \bibinfo {author} {\bibfnamefont
  {W.}~\bibnamefont {Hanke}}, \bibinfo {author} {\bibfnamefont
  {R.}~\bibnamefont {Thomale}}, \bibinfo {author} {\bibfnamefont
  {J.}~\bibnamefont {Sch{\"a}fer}}, \ and\ \bibinfo {author} {\bibfnamefont
  {R.}~\bibnamefont {Claessen}},\ }\bibfield  {title} {\enquote {\bibinfo
  {title} {Bismuthene on a {SiC} substrate: {A} candidate for a
  high-temperature quantum spin {Hall} material},}\ }\href {\doibase
  10.1126/science.aai8142} {\bibfield  {journal} {\bibinfo  {journal}
  {Science}\ }\textbf {\bibinfo {volume} {357}},\ \bibinfo {pages} {287}
  (\bibinfo {year} {2017})}\BibitemShut {NoStop}%
\bibitem [{\citenamefont {Lu}\ \emph {et~al.}(2016)\citenamefont {Lu},
  \citenamefont {Zhou}, \citenamefont {Chang}, \citenamefont {Guan},
  \citenamefont {Chen}, \citenamefont {Jiang}, \citenamefont {Jiang},
  \citenamefont {Wang}, \citenamefont {Yang},\ and\ \citenamefont
  {Feng}}]{lu2016multiple}%
  \BibitemOpen
  \bibfield  {author} {\bibinfo {author} {\bibfnamefont {Y.}~\bibnamefont
  {Lu}}, \bibinfo {author} {\bibfnamefont {D.}~\bibnamefont {Zhou}}, \bibinfo
  {author} {\bibfnamefont {G.}~\bibnamefont {Chang}}, \bibinfo {author}
  {\bibfnamefont {S.}~\bibnamefont {Guan}}, \bibinfo {author} {\bibfnamefont
  {W.}~\bibnamefont {Chen}}, \bibinfo {author} {\bibfnamefont {Y.}~\bibnamefont
  {Jiang}}, \bibinfo {author} {\bibfnamefont {J.}~\bibnamefont {Jiang}},
  \bibinfo {author} {\bibfnamefont {X.}~\bibnamefont {Wang}}, \bibinfo {author}
  {\bibfnamefont {S.~A.}\ \bibnamefont {Yang}}, \ and\ \bibinfo {author}
  {\bibfnamefont {Y.~P.}\ \bibnamefont {Feng}},\ }\bibfield  {title} {\enquote
  {\bibinfo {title} {Multiple unpinned {Dirac} points in group-{V}a
  single-layers with phosphorene structure},}\ }\href {\doibase
  10.1038/npjcompumats.2016.11;} {\bibfield  {journal} {\bibinfo  {journal}
  {npj Comput. Mater.}\ }\textbf {\bibinfo {volume} {2}},\ \bibinfo {pages}
  {16011} (\bibinfo {year} {2016})}\BibitemShut {NoStop}%
\bibitem [{\citenamefont {Maslov}\ \emph {et~al.}(2022)\citenamefont {Maslov},
  \citenamefont {Kumar},\ and\ \citenamefont {Maiti}}]{maslov2022collective}%
  \BibitemOpen
  \bibfield  {author} {\bibinfo {author} {\bibfnamefont {D.~L.}\ \bibnamefont
  {Maslov}}, \bibinfo {author} {\bibfnamefont {A.}~\bibnamefont {Kumar}}, \
  and\ \bibinfo {author} {\bibfnamefont {S.}~\bibnamefont {Maiti}},\ }\bibfield
   {title} {\enquote {\bibinfo {title} {{Collective} {Spin} {Modes} in {Fermi}
  {Liquids} with {Spin}-{Orbit} {Coupling}},}\ }\href {\doibase
  10.1134/S1063776122100077} {\bibfield  {journal} {\bibinfo  {journal} {JETP}\
  }\textbf {\bibinfo {volume} {135}},\ \bibinfo {pages} {549} (\bibinfo {year}
  {2022})}\BibitemShut {NoStop}%
\bibitem [{\citenamefont {Choi}\ \emph {et~al.}(2023)\citenamefont {Choi},
  \citenamefont {Lane}, \citenamefont {Zhu},\ and\ \citenamefont
  {Crooker}}]{Choi2023:NM}%
  \BibitemOpen
  \bibfield  {author} {\bibinfo {author} {\bibfnamefont {J.}~\bibnamefont
  {Choi}}, \bibinfo {author} {\bibfnamefont {C.}~\bibnamefont {Lane}}, \bibinfo
  {author} {\bibfnamefont {J.-X.}\ \bibnamefont {Zhu}}, \ and\ \bibinfo
  {author} {\bibfnamefont {S.~A.}\ \bibnamefont {Crooker}},\ }\bibfield
  {title} {\enquote {\bibinfo {title} {{Asymmetric magnetic proximity
  interactions in MoSe$_2$/CrBr$_3$ van der Waals heterostructures}},}\ }\href
  {\doibase 10.1038/s41563-022-01424-w} {\bibfield  {journal} {\bibinfo
  {journal} {Nat. Mater.}\ }\textbf {\bibinfo {volume} {22}},\ \bibinfo {pages}
  {305} (\bibinfo {year} {2023})}\BibitemShut {NoStop}%
\bibitem [{\citenamefont {Zhou}\ and\ \citenamefont
  {\v{Z}ut\'c}(2023)}]{Zhou2023:NM}%
  \BibitemOpen
  \bibfield  {author} {\bibinfo {author} {\bibfnamefont {T.}~\bibnamefont
  {Zhou}}\ and\ \bibinfo {author} {\bibfnamefont {I.}~\bibnamefont
  {\v{Z}ut\'c}},\ }\bibfield  {title} {\enquote {\bibinfo {title} {Asymmetry in
  the magnetic neighbourhood},}\ }\href {\doibase 10.1038/s41563-022-01466-0}
  {\bibfield  {journal} {\bibinfo  {journal} {Nat. Mater.}\ }\textbf {\bibinfo
  {volume} {22}},\ \bibinfo {pages} {284} (\bibinfo {year} {2023})}\BibitemShut
  {NoStop}%
\bibitem [{\citenamefont {Ganichev}\ \emph {et~al.}(2002)\citenamefont
  {Ganichev}, \citenamefont {Ivchenko}, \citenamefont {Bel'Kov}, \citenamefont
  {Tarasenko}, \citenamefont {Sollinger}, \citenamefont {Weiss}, \citenamefont
  {Wegscheider},\ and\ \citenamefont {Prettl}}]{Ganichev2002:N}%
  \BibitemOpen
  \bibfield  {author} {\bibinfo {author} {\bibfnamefont {S.~D.}\ \bibnamefont
  {Ganichev}}, \bibinfo {author} {\bibfnamefont {E.~L.}\ \bibnamefont
  {Ivchenko}}, \bibinfo {author} {\bibfnamefont {V.~V.}\ \bibnamefont
  {Bel'Kov}}, \bibinfo {author} {\bibfnamefont {S.~A.}\ \bibnamefont
  {Tarasenko}}, \bibinfo {author} {\bibfnamefont {M.}~\bibnamefont
  {Sollinger}}, \bibinfo {author} {\bibfnamefont {D.}~\bibnamefont {Weiss}},
  \bibinfo {author} {\bibfnamefont {W.}~\bibnamefont {Wegscheider}}, \ and\
  \bibinfo {author} {\bibfnamefont {W.}~\bibnamefont {Prettl}},\ }\bibfield
  {title} {\enquote {\bibinfo {title} {Spin-galvanic effect},}\ }\href@noop {}
  {\bibfield  {journal} {\bibinfo  {journal} {Nature}\ }\textbf {\bibinfo
  {volume} {417}},\ \bibinfo {pages} {153} (\bibinfo {year}
  {2002})}\BibitemShut {NoStop}%
\bibitem [{\citenamefont {Yang}\ \emph {et~al.}(2022)\citenamefont {Yang},
  \citenamefont {Valenzuela}, \citenamefont {Chshiev}, \citenamefont {Couet},
  \citenamefont {Dieny}, \citenamefont {Dlubak}, \citenamefont {Fert},
  \citenamefont {Garello}, \citenamefont {Jamet}, \citenamefont {Jeong},\ and\
  \citenamefont {et~al.}}]{Yang2022:N}%
  \BibitemOpen
  \bibfield  {author} {\bibinfo {author} {\bibfnamefont {H.}~\bibnamefont
  {Yang}}, \bibinfo {author} {\bibfnamefont {S.~O.}\ \bibnamefont
  {Valenzuela}}, \bibinfo {author} {\bibfnamefont {M.}~\bibnamefont {Chshiev}},
  \bibinfo {author} {\bibfnamefont {S.}~\bibnamefont {Couet}}, \bibinfo
  {author} {\bibfnamefont {B.}~\bibnamefont {Dieny}}, \bibinfo {author}
  {\bibfnamefont {B.}~\bibnamefont {Dlubak}}, \bibinfo {author} {\bibfnamefont
  {A.}~\bibnamefont {Fert}}, \bibinfo {author} {\bibfnamefont {K.}~\bibnamefont
  {Garello}}, \bibinfo {author} {\bibfnamefont {M.}~\bibnamefont {Jamet}},
  \bibinfo {author} {\bibfnamefont {D.-E.}\ \bibnamefont {Jeong}}, \ and\
  \bibinfo {author} {\bibnamefont {et~al.}},\ }\bibfield  {title} {\enquote
  {\bibinfo {title} {Two-dimensional materials prospects for non-volatile
  spintronic memories},}\ }\href@noop {} {\bibfield  {journal} {\bibinfo
  {journal} {Nature}\ }\textbf {\bibinfo {volume} {606}},\ \bibinfo {pages}
  {663} (\bibinfo {year} {2022})}\BibitemShut {NoStop}%
\bibitem [{\citenamefont {Dainone}\ \emph {et~al.}(2024)\citenamefont
  {Dainone}, \citenamefont {Renucci}, \citenamefont {Bouche}, \citenamefont
  {Prestes}, \citenamefont {Morassi}, \citenamefont {Devaux}, \citenamefont
  {Lindemann}, \citenamefont {George}, \citenamefont {Jaffres}, \citenamefont
  {Lemaitre},\ and\ \citenamefont {et~al.}}]{Dainone2024:N}%
  \BibitemOpen
  \bibfield  {author} {\bibinfo {author} {\bibfnamefont {P.~A.}\ \bibnamefont
  {Dainone}}, \bibinfo {author} {\bibfnamefont {P.}~\bibnamefont {Renucci}},
  \bibinfo {author} {\bibfnamefont {A.}~\bibnamefont {Bouche}}, \bibinfo
  {author} {\bibfnamefont {N.~F.}\ \bibnamefont {Prestes}}, \bibinfo {author}
  {\bibfnamefont {M.}~\bibnamefont {Morassi}}, \bibinfo {author} {\bibfnamefont
  {X.}~\bibnamefont {Devaux}}, \bibinfo {author} {\bibfnamefont
  {M.}~\bibnamefont {Lindemann}}, \bibinfo {author} {\bibfnamefont {J.-M.}\
  \bibnamefont {George}}, \bibinfo {author} {\bibfnamefont {H.}~\bibnamefont
  {Jaffres}}, \bibinfo {author} {\bibfnamefont {A.}~\bibnamefont {Lemaitre}}, \
  and\ \bibinfo {author} {\bibnamefont {et~al.}},\ }\bibfield  {title}
  {\enquote {\bibinfo {title} {Controlling the helicity of light by electrical
  magnetization switching},}\ }\href@noop {} {\bibfield  {journal} {\bibinfo
  {journal} {Nature}\ }\textbf {\bibinfo {volume} {in press}} (\bibinfo {year}
  {2024})}\BibitemShut {NoStop}%
\bibitem [{\citenamefont {Tsymbal}\ and\ \citenamefont
  {{\v{Z}}uti{\'c}}(2012)}]{Tsymbal:2019}%
  \BibitemOpen
  \bibfield  {author} {\bibinfo {author} {\bibfnamefont {E.~Y.}\ \bibnamefont
  {Tsymbal}}\ and\ \bibinfo {author} {\bibfnamefont {I.}~\bibnamefont
  {{\v{Z}}uti{\'c}}},\ }\href@noop {} {\emph {\bibinfo {title} {Spintronics
  Handbook Spin Transport and Magnetism, 2nd Edition}}}\ (\bibinfo  {publisher}
  {CRC Press, Taylor and Francis, Boca Raton, FL},\ \bibinfo {year}
  {2012})\BibitemShut {NoStop}%
\bibitem [{\citenamefont {Jhuria}\ \emph {et~al.}(2020)\citenamefont {Jhuria},
  \citenamefont {Hohlfeld}, \citenamefont {Pattabi}, \citenamefont {Martin},
  \citenamefont {Arriola~C{\'o}rdova}, \citenamefont {Shi}, \citenamefont
  {Lo~Conte}, \citenamefont {Petit-Watelot}, \citenamefont {Rojas-Sanchez},
  \citenamefont {Malinowski},\ and\ \citenamefont {et~al.}}]{Jhuria2020:NE}%
  \BibitemOpen
  \bibfield  {author} {\bibinfo {author} {\bibfnamefont {K.}~\bibnamefont
  {Jhuria}}, \bibinfo {author} {\bibfnamefont {J.}~\bibnamefont {Hohlfeld}},
  \bibinfo {author} {\bibfnamefont {A.}~\bibnamefont {Pattabi}}, \bibinfo
  {author} {\bibfnamefont {E.}~\bibnamefont {Martin}}, \bibinfo {author}
  {\bibfnamefont {A.~Y.}\ \bibnamefont {Arriola~C{\'o}rdova}}, \bibinfo
  {author} {\bibfnamefont {X.}~\bibnamefont {Shi}}, \bibinfo {author}
  {\bibfnamefont {R.}~\bibnamefont {Lo~Conte}}, \bibinfo {author}
  {\bibfnamefont {S.}~\bibnamefont {Petit-Watelot}}, \bibinfo {author}
  {\bibfnamefont {J.~C.}\ \bibnamefont {Rojas-Sanchez}}, \bibinfo {author}
  {\bibfnamefont {G.}~\bibnamefont {Malinowski}}, \ and\ \bibinfo {author}
  {\bibnamefont {et~al.}},\ }\bibfield  {title} {\enquote {\bibinfo {title}
  {Spin-orbit torque switching of a ferromagnet with picosecond electrical
  pulses},}\ }\href@noop {} {\bibfield  {journal} {\bibinfo  {journal} {Nat.
  Electron.}\ }\textbf {\bibinfo {volume} {3}},\ \bibinfo {pages} {680}
  (\bibinfo {year} {2020})}\BibitemShut {NoStop}%
\bibitem [{\citenamefont {Sodemann}\ and\ \citenamefont
  {Fu}(2015)}]{sodemann2015quantum}%
  \BibitemOpen
  \bibfield  {author} {\bibinfo {author} {\bibfnamefont {I.}~\bibnamefont
  {Sodemann}}\ and\ \bibinfo {author} {\bibfnamefont {L.}~\bibnamefont {Fu}},\
  }\bibfield  {title} {\enquote {\bibinfo {title} {Quantum nonlinear {Hall}
  effect induced by {Berry} curvature dipole in time-reversal invariant
  materials},}\ }\href {\doibase 10.1103/PhysRevLett.115.216806} {\bibfield
  {journal} {\bibinfo  {journal} {Phys. Rev. Lett.}\ }\textbf {\bibinfo
  {volume} {115}},\ \bibinfo {pages} {216806} (\bibinfo {year}
  {2015})}\BibitemShut {NoStop}%
\bibitem [{\citenamefont {Malla}\ \emph {et~al.}(2021)\citenamefont {Malla},
  \citenamefont {Saxena},\ and\ \citenamefont {Kort-Kamp}}]{malla2021emerging}%
  \BibitemOpen
  \bibfield  {author} {\bibinfo {author} {\bibfnamefont {R.~K.}\ \bibnamefont
  {Malla}}, \bibinfo {author} {\bibfnamefont {A.}~\bibnamefont {Saxena}}, \
  and\ \bibinfo {author} {\bibfnamefont {W.~J.~M.}\ \bibnamefont {Kort-Kamp}},\
  }\bibfield  {title} {\enquote {\bibinfo {title} {Emerging nonlinear {Hall}
  effect in {Kane-Mele} two-dimensional topological insulators},}\ }\href
  {\doibase 10.1103/PhysRevB.104.205422} {\bibfield  {journal} {\bibinfo
  {journal} {Phys. Rev. B}\ }\textbf {\bibinfo {volume} {104}},\ \bibinfo
  {pages} {205422} (\bibinfo {year} {2021})}\BibitemShut {NoStop}%
\bibitem [{\citenamefont {Denisov}(2022)}]{denisov2022electric}%
  \BibitemOpen
  \bibfield  {author} {\bibinfo {author} {\bibfnamefont {K.~S.}\ \bibnamefont
  {Denisov}},\ }\bibfield  {title} {\enquote {\bibinfo {title} {Electric
  {field} {effect} on {electron} {gas} {spins} in {two}-{dimensional} {magnets}
  with {strong} {spin}-{orbit} {coupling}},}\ }\href {\doibase
  10.1103/PhysRevB.105.045413} {\bibfield  {journal} {\bibinfo  {journal}
  {Phys. Rev. B}\ }\textbf {\bibinfo {volume} {105}},\ \bibinfo {pages}
  {045413} (\bibinfo {year} {2022})}\BibitemShut {NoStop}%
\bibitem [{\citenamefont {Yao}\ \emph {et~al.}(2004)\citenamefont {Yao},
  \citenamefont {Kleinman}, \citenamefont {MacDonald}, \citenamefont {Sinova},
  \citenamefont {Jungwirth}, \citenamefont {Wang}, \citenamefont {Wang},\ and\
  \citenamefont {Niu}}]{yao2004first}%
  \BibitemOpen
  \bibfield  {author} {\bibinfo {author} {\bibfnamefont {Y.}~\bibnamefont
  {Yao}}, \bibinfo {author} {\bibfnamefont {L.}~\bibnamefont {Kleinman}},
  \bibinfo {author} {\bibfnamefont {A.~H.}\ \bibnamefont {MacDonald}}, \bibinfo
  {author} {\bibfnamefont {J.}~\bibnamefont {Sinova}}, \bibinfo {author}
  {\bibfnamefont {T.}~\bibnamefont {Jungwirth}}, \bibinfo {author}
  {\bibfnamefont {D.}~\bibnamefont {Wang}}, \bibinfo {author} {\bibfnamefont
  {E.}~\bibnamefont {Wang}}, \ and\ \bibinfo {author} {\bibfnamefont
  {Q.}~\bibnamefont {Niu}},\ }\bibfield  {title} {\enquote {\bibinfo {title}
  {First {Principles} {Calculation} of {Anomalous} $\mathrm{Hall}$
  {Conductivity} in {Ferromagnetic} bcc $\mathrm{Fe}$},}\ }\href {\doibase
  10.1103/PhysRevLett.92.037204} {\bibfield  {journal} {\bibinfo  {journal}
  {Phys. Rev. Lett.}\ }\textbf {\bibinfo {volume} {92}},\ \bibinfo {pages}
  {037204} (\bibinfo {year} {2004})}\BibitemShut {NoStop}%
\bibitem [{\citenamefont {Bray}\ \emph {et~al.}(2022)\citenamefont {Bray},
  \citenamefont {Maussang}, \citenamefont {Consejo}, \citenamefont
  {Delgado-Notario}, \citenamefont {Krishtopenko}, \citenamefont {Yahniuk},
  \citenamefont {Gebert}, \citenamefont {Ruffenach}, \citenamefont {Dinar},
  \citenamefont {Moench},\ and\ \citenamefont {et~al.}}]{PhysRevB.106.245141}%
  \BibitemOpen
  \bibfield  {author} {\bibinfo {author} {\bibfnamefont {C.}~\bibnamefont
  {Bray}}, \bibinfo {author} {\bibfnamefont {K.}~\bibnamefont {Maussang}},
  \bibinfo {author} {\bibfnamefont {C.}~\bibnamefont {Consejo}}, \bibinfo
  {author} {\bibfnamefont {J.~A.}\ \bibnamefont {Delgado-Notario}}, \bibinfo
  {author} {\bibfnamefont {S.}~\bibnamefont {Krishtopenko}}, \bibinfo {author}
  {\bibfnamefont {I.}~\bibnamefont {Yahniuk}}, \bibinfo {author} {\bibfnamefont
  {S.}~\bibnamefont {Gebert}}, \bibinfo {author} {\bibfnamefont
  {S.}~\bibnamefont {Ruffenach}}, \bibinfo {author} {\bibfnamefont
  {K.}~\bibnamefont {Dinar}}, \bibinfo {author} {\bibfnamefont
  {E.}~\bibnamefont {Moench}}, \ and\ \bibinfo {author} {\bibnamefont
  {et~al.}},\ }\bibfield  {title} {\enquote {\bibinfo {title}
  {Temperature-{dependent} {zero}-{field} {splittings} in {graphene}},}\ }\href
  {\doibase 10.1103/PhysRevB.106.245141} {\bibfield  {journal} {\bibinfo
  {journal} {Phys. Rev. B}\ }\textbf {\bibinfo {volume} {106}},\ \bibinfo
  {pages} {245141} (\bibinfo {year} {2022})}\BibitemShut {NoStop}%
\end{thebibliography}%
%EDSR_Ref

\end{document}